\documentclass[preprint2]{aastex61}

\renewcommand\plotone[1]{\includegraphics[width=\linewidth]{#1}}

\newcommand\pirel{\pi_{\rm rel}}
\newcommand\piabs{\pi_{\rm abs}}
\newcommand\murel{\mu_{\rm rel}}
\newcommand\mas{{\rm mas}}
\newcommand\masyear{{\rm mas~yr^{-1}}}
\newcommand\kmsec{{\rm km~s^{-1}}}
\newcommand\vtan{V_{\rm tan}}
\newcommand\pa{{\rm P.~A.~}}
                                                 
\shorttitle{CCD Parallaxes}
\shortauthors{Dahn et al.}

\begin{document}

\title{CCD Parallaxes for 309 Late-type Dwarfs and Subdwarfs}

\author{Conard C. Dahn}
\affiliation{US Naval Observatory, Flagstaff Station, 10391 W. Naval Observatory
Road, Flagstaff, AZ 86005-8521, USA}

\author{Hugh C. Harris}
\affiliation{US Naval Observatory, Flagstaff Station, 10391 W. Naval Observatory
Road, Flagstaff, AZ 86005-8521, USA}

\author{John P. Subasavage}
\affiliation{US Naval Observatory, Flagstaff Station, 10391 W. Naval Observatory
Road, Flagstaff, AZ 86005-8521, USA}

\author{Harold D. Ables}
\affiliation{US Naval Observatory, Flagstaff Station, 10391 W. Naval Observatory
Road, Flagstaff, AZ 86005-8521, USA}

\author{Blaise J. Canzian}
\affiliation{L-3 Communications/Brashear, 615 Epsilon Dr., Pittsburgh, PA 15238-2807, USA}

\author{Harry H. Guetter}
\affiliation{US Naval Observatory, Flagstaff Station, 10391 W. Naval Observatory
Road, Flagstaff, AZ 86005-8521, USA}

\author{Fred H. Harris}
\affiliation{US Naval Observatory, Flagstaff Station, 10391 W. Naval Observatory
Road, Flagstaff, AZ 86005-8521, USA}

\author{Arne H. Henden}
\affiliation{AAVSO, Cambridge, MA 02138, USA}

\author{S. K. Leggett}
\affiliation{Gemini Observatory, Northern Operations Center, 670 N. A'ohoku Place, Hilo, HI 96720, USA}

\author{Stephen E. Levine}
\affiliation{Lowell Observatory, 1400 W. Mars Hill Road, Flagstaff, AZ  86001-4499, USA}

\author{Christian B. Luginbuhl}
\affiliation{US Naval Observatory, Flagstaff Station, 10391 W. Naval Observatory
Road, Flagstaff, AZ 86005-8521, USA}

\author{Alice B. Monet}
\affiliation{US Naval Observatory, Flagstaff Station, 10391 W. Naval Observatory
Road, Flagstaff, AZ 86005-8521, USA}

\author{David G. Monet}
\affiliation{US Naval Observatory, Flagstaff Station, 10391 W. Naval Observatory
Road, Flagstaff, AZ 86005-8521, USA}

\author{Jeffrey A. Munn}
\affiliation{US Naval Observatory, Flagstaff Station, 10391 W. Naval Observatory
Road, Flagstaff, AZ 86005-8521, USA}

\author{Jeffrey R. Pier}
\affiliation{US Naval Observatory, Flagstaff Station, 10391 W. Naval Observatory
Road, Flagstaff, AZ 86005-8521, USA}

\author{Ronald C. Stone}
\altaffiliation{Deceased}
\affiliation{US Naval Observatory, Flagstaff Station, 10391 W. Naval Observatory
Road, Flagstaff, AZ 86005-8521, USA}

\author{Frederick J. Vrba}
\affiliation{US Naval Observatory, Flagstaff Station, 10391 W. Naval Observatory
Road, Flagstaff, AZ 86005-8521, USA}

\author{Richard L. Walker}
\altaffiliation{Deceased}
\affiliation{US Naval Observatory, Flagstaff Station, 10391 W. Naval Observatory
Road, Flagstaff, AZ 86005-8521, USA}

\author{Trudy M. Tilleman}
\affiliation{US Naval Observatory, Flagstaff Station, 10391 W. Naval Observatory
Road, Flagstaff, AZ 86005-8521, USA}

\correspondingauthor{Subasavage, J. P.}
\email{jsubasavage@nofs.navy.mil}

\begin{abstract}
New, updated, and/or revised CCD parallaxes determined with the Strand
Astrometric Reflector at the Naval Observatory Flagstaff Station (NOFS)
are presented.  Included are results for 309 late-type dwarf and subdwarf
stars observed over the 30+ years that the program operated.  For 124 of
the stars, parallax determinations from other investigators have already
appeared in the literature and we compare the different results.  Also included
here is new or updated $VI$ photometry on the Johnson-Kron-Cousins system for all
but a few of the faintest targets.  Together with 2MASS $JHK_s$ near-infrared
photometry, a sample of absolute magnitude versus color and
color versus color diagrams are constructed.  Since large proper motion was
a prime criterion for targeting the stars, the majority turn out to be either
M-type subdwarfs or late M-type dwarfs.  The sample also includes 50 dwarf
or subdwarf L-type stars, and four T dwarfs.  Possible halo subdwarfs are identified in the sample based on tangential velocity, subluminosity, and spectral type. Residuals from the solutions for
parallax and proper motion for several stars show evidence of astrometric
perturbations.
\end{abstract}


\section{Introduction}
\explain{The referee's suggested changes in units (deg to $^\circ$, yrs to yr, sec to s; his suggestions numbered 9, 11, 12) have all been implemented, however they have not been highlighted with change tracking, due to the numerous changes and their trivial nature.  Similarly, the changing figure and paneling numbering is not highlighted in the text.  The remaining changes have all been indicated with change tracking}.
Ground-based trigonometric parallax observations have provided the anchor for the cosmological distance ladder for over a \replaced{centutry}{century}.  Such parallax determinations were primarily made
photographically using classical refracting telescopes with their
associated accidental and systematic errors.  \citet{1995gcts.book.....V}
presented a compilation (hereafter referred to as YPC4) of all trigonometric stellar parallaxes
available through 1995 November -- 15,994 determinations for a total
of 8112 individual stars.  Included is a thorough analysis of the
associated errors for each telescope employed.  During the late 1980s
improved precisions were realized by using hypersensitized
high-DQE photographic emulsions and more sophisticated plate measuring
machines.  However, parallax determinations with formal mean errors
below $\sim$ 2~mas were rarely obtained photographically.

The U.S. Naval Observatory 61-in Astrometric Reflector \citep{1964S&T....27..204S} was
commissioned at the Flagstaff Station in late 1963 and commenced
regular observations for stellar trigonometric parallaxes in mid 1964.
The Strand Astrometric Reflector was designed specifically for
high-precision narrow-field differential astrometric observations.
Optically, it consists of a parabolic primary mirror along with a flat
secondary mirror and arranged in a Cassegrain
configuration.  With no optical power from the secondary mirror, the
main optical aberration is coma from the primary mirror.  Only in
exceptionally good seeing was the coma detectable within the field
generally employed for photographic parallax observations on
$5 \times 7$-inch Kodak plates.  Furthermore, since the focal plane is flat and
the distortion produced by coma is radially symmetric around the
optical axis, it is allowed for by the scale terms in the linear plate
constant solutions employed.  On the negative side, such an optical
configuration necessitates a large secondary mirror (with cell,
approximately 40 inch diameter) in order to produce an unvignetted field
of $\sim30$~arcmin in radius.  This effectively reduces the light-gathering
power of the telescope to one of $\sim 49$ inch clear aperture.  The effective
focal plane scale of the Strand Reflector has been monitored both with
time and position across the FOV employed.  To four significant
digits, it is constant at 13.55 arcsec~mm$^{-1}$.

Operating in photographic mode through mid 1994, a total of 1013 Naval Observatory Flagstaff Station (NOFS)
parallaxes were derived.  While parallax determinations with \replaced{formal mean errors}{uncertainties} in the 2 -- 4 mas range were routinely obtained,
precisions in the 1 -- 2 mas range were still rarely realized.
Speculation grew that this formal error ``barrier'' might be either due to
either fundamentals associated with making ground-based observations
through the Earth's turbulent atmosphere \citep{1980A+A....89...41L,1987AJ.....94..213G,1989AJ.....97..607H} and/or due to slight emulsion shifts perhaps
associated with plate processing \citep{1977AJ.....82..299L}.  However,
when the NOFS parallax program transitioned from photography to a CCD
detector in the mid 1980s, it soon became apparent that
parallax \replaced{errors}{uncertainties} at the 0.5 -- 1.2 mas level could routinely be
realized.

In this paper, we present a compilation of NOFS CCD parallaxes for all
309 of the late-type dwarf and subdwarf stars that have been observed
with the Strand Astrometric Telescope.  The results presented here
include all observations obtained through 2016 June and employ all of
the processing updates outlined in the following section.  A similar
compilation for $\sim$ 170 white dwarf stars will be presented in a separate
paper, which is in preparation.  Not included in this paper are the results
for the three perturbation binaries for which preliminary results were
presented in \citet{2015csss...18..413H} and which are under continuing
observation.  NOFS results for the perturbation binary LSR1610-00 have
recently been presented by \citet{2016AJ....151...57K} and will not be
repeated here.  Also not included here are results for 10 dwarf carbon
stars being examined for astrometric perturbations with continuing
observations.

Section 2 summarizes the evolution of the NOFS CCD parallax work,
including various refinements and improvements -- both in
instrumentation and data processing -- which have been instituted over
the 25 year duration of the program.  The final astrometric results
are presented in Section 3 and the supporting photometric results for
the target stars are given in Section 4.   Section 5 presents several color-magnitude and
color-color diagrams constructed from our database.  An additional discussion regarding anticipated {\it  Gaia} \citep{2016A&A...595A...2G, 2016A&A...595A...1G} contributions for late-type dwarfs and subdwarfs follows in Section 6.

\section{Evolution of the NOFS CCD Parallax Program}
\label{sec:evolution}

By way of background, we note that the NOFS CCD parallax program
evolved from the efforts initiated by D. Monet using the KPNO 4m/CCD
instrumentation \citep{1983AJ.....88.1489M}.  The history of and details on
the early USNO efforts can be found in \citet{1992AJ....103..638M} where
results are presented for 72 stars.  The TI800 camera employed in
this early work was very crude by modern standards in that the chip
was cosmetically poor, having several blocked columns, and, at the
Strand Telescope, provided a limited useful FOV of only $2.5 \times 2.5$
arcminutes.  On the positive side, the small pixel size (0.206\arcsec\ on the
sky) provided excellent image sampling, even in the very best seeing
conditions realized at the Strand Telescope ($\sim$~0.5\arcsec\ FWHM).

As noted in \citet[][Table 1]{1992AJ....103..638M}, observations continued on the
majority of the fields following publication of these preliminary
results.  Starting in late 1992, a camera employing a thinned,
backside-illuminated Tektronix $2048 \times 2048$ pixel CCD, with a flatness
characterized by a raised hump of not more than 220 $\micron$
peak to valley, was tested and
commissioned.  This so-called TEK2K camera provides a generous $11 \times 11$
arcmin FOV and, hence, a much richer selection of potential reference
stars for the differential astrometric measures.  Although the image
sampling is somewhat degraded (24.0 $\micron$ pixels giving
0.325\arcsec\ pixel$^{-1}$), good centroids were obtained.  Consequently, the TEK2K
camera rapidly became the workhorse for the NOFS parallax observations
and the TI800 camera was formally decommissioned in 1995.  The
observational procedures employed with the TEK2K camera have been
summarized in \citet[][Section 2.1]{2002AJ....124.1170D}.

After only a few years of experience using the TEK2K camera, it became
apparent that relative parallaxes with \replaced{formal mean errors}{uncertainties} $\sim0.5$~mas
were routinely being achieved.  This improvement can be attributed
primarily to the much better reference star frames available using
this larger FOV CCD.  However, along with the improved relative
parallaxes came the need for higher quality determinations for the
correction from relative to absolute parallax -- the quantity
providing the actual distance determinations needed for astrophysical
applications.  (For additional information regarding this correction,
see the footnote on page 1173 of \citealt{2002AJ....124.1170D}).  For the TI800
results presented in \citet{1992AJ....103..638M}, we -- following YPC4 --
employed their method of using the average magnitude of the reference
stars employed in the field of each individual parallax target along
with a galactic model to estimate the average reference star frame
distance for each galactic direction.  Although this approach was
entirely adequate for the parallax determinations in YPC4, applying it
for the higher precision TEK2K relative parallaxes would significantly
inflate the \replaced{formal error}{uncertainties} derived for the absolute parallaxes.  Since
we were already measuring $V$, $V-I$ photometry of each individual
reference star employed in each parallax target field to calculate
corrections for differential color refraction \citep[DCR; see][Section 3.2]{1992AJ....103..638M}, we adopted a calibrated version of the $M_V$ versus
$V-I$ color-magnitude diagram to determine photometric parallaxes for
each star.  (For further details on this calibration, see \citealt{2016AJ....152..118H}.)  Photometry from the 2MASS Point-Source Catalog
\citep{2006AJ....131.1163S} is employed, when required, to resolve
ambiguities in dwarf versus giant status of reference stars when the
relative astrometry is not decisive and to clarify issues regarding
interstellar extinction/reddening in fields at low galactic latitudes.

The higher precision TEK2K relative parallax determinations also
necessitated an upgrade in the determination of parallax factors --
the projected baseline along the Earth-Sun direction at each epoch of
observation.  These quantities enter the equations of conditions used
in the least-squares solution for the parallax as multiplicative
products with the parallax.  Since the NOFS parallax determinations
are mainly for stars at distances between 10 and 300 parsecs, parallax
factors calculated to four significant figures are required.  Hence,
values calculated for the projected distance between the Earth and the
solar system barycenter were adopted.

A third camera employing an EEV (English Electric Valve; now e2v)
$2048 \times 4096$ pixel CCD42-80 thinned backside-illuminated CCD was
commissioned in early 2008.  This device has
13.5 $\micron$ pixels, corresponding to 0.183\arcsec\ ~pixel$^{-1}$ on the sky.
Hence, it provides better image sampling than TEK2K and gives an
FOV of $6.2 \times 12.4$ arcmin on the sky.  Furthermore, the pixels of this
device have larger full-well charge capacity than the TEK2K chip, and
this results in a slightly enhanced dynamic range before saturation
effects become important.
This detector had suffered a hygroscopic exposure to its
anti-reflection (AR) coating while in use in tests supporting the
ill-fated USNO Full-Sky Astrometric Mapping Explorer satellite engineering effort, resulting in discoloration of
the backside surface when visibly inspected.  The AR coating mostly
returned to its manufactured state after six months of storage in a
vacuum environment.  The CCD's deviation from flatness is not more than 15 $\micron$
peak to valley.  The scheduled observing time on the Strand
reflector around new moon then became shared between TEK2K and EEV24.

Another camera, employing an engineering grade Tektronix $2048 \times 2048$ CCD
with a circular ($\sim3$~mm diameter, $\sim40$~arcsec on the sky), highly uniform
neutral density attenuation spot (Inconel on an optically flat quartz
substrate), mounted $\sim1$~mm in front of the CCD, was commissioned in
late 1995.  The spot provides $\sim9$.0 magnitudes of brightness reduction
for the parallax target star, permitting targets as bright as fourth
magnitude to be measured.  The original motivation for this camera --
designated ND9 -- was concern about possible degradation in the
astrometry from the ESA {|it Hipparcos} satellite due its failure of the
craft to reach its intended geostationary orbit following launch in
1989 August.  Happily, the final {\it Hipparcos} parallaxes obtained from
the 4 year mission proved to be determined to the 1~mas level, as
planned.  Instead, a limited program was initiated with ND9 for the purpose of
demonstrating that precision astrometry could be achieved from the
ground using this form of magnitude compensation.  68 {\it Hipparcos} stars
were observed over an epoch range of 5.2 years (on average) during the
period from 1996 to 2002.  A sample of preliminary results were presented in \citet{1997ESASP.402..105H}.  The derived ND9 parallaxes agreed well with the
published values from {\it Hipparcos}, with \replaced{formal parallax errors}{parallax uncertainties}
generally better than their {\it Hipparcos} counterparts, except in a few
cases.  ND9 was decommissioned from routine parallax work in early
2002.
\section{Astrometric Results}

We continue to employ the image centroiding method developed by
D. Monet \citep[cf.][Section III.b]{1983AJ.....88.1489M} which employs an integral
of the illumination under a circular symmetric Gaussian as the
algorithm.  Head-to-head comparison with other techniques employed by
DAOPHOT \citep{1987PASP...99..191S} and SExtractor \citep{1996A&AS..117..393B} show that the
Monet method is at least as stable in uncrowded fields where
uncontaminated images are encountered.  In fields where the target star is blended with neighboring objects, DAOPHOT was used for deblending and centroiding.  However, the Monet method is not appropriate
for photometric reductions and is not employed for those observations
(see Section~\ref{section:photometry} below).

Our solutions for parallax are carried out independently in both RA
and DEC.  These values are then combined as a weighted average to
obtain the final value presented below.  The validity of this
approach was discussed by \citet{1980AJ.....85.1390L}, and offers the
advantage of identifying bad data (e.g., a cosmic-ray strike in the
wings of an image) and providing an alert to possible astrometric
perturbations.

The astrometric results are presented in Table~\ref{tab:ast}. These determinations
include all acceptable observations obtained through 2016 June and
employ all of the reduction improvements outlined in the previous
section.  Results for a total of 309 individual stars are tabulated
with two independent determinations included for four stars (LP 540-16,
LHS 2397a, LHS 3406, and LHS 474).  Column (1) in Table~\ref{tab:ast} gives the
identifying number in the 2MASS Point Source Catalog \citep{2006AJ....131.1163S}.  The 2MASS J number provides an unambiguous link to SIMBAD
where many alternate names and much additional information can be
found.  A few stars are not individually found in 2MASS, usually due to
their proximity to another star at the epochs of the 2MASS
observations.  For these stars, we adopt their LHS designation in column (2), and their coordinates from {\it Gaia} DR1 \citep{2016A+A...595A...4L}.

\begin{splitdeluxetable*}{cccccccccccBcr@{$\pm$}rr@{$\pm$}rr@{$\pm$}rr@{$\pm$}rr@{$\pm$}rl}
\voffset-10pt{}
\tabletypesize{\tiny}
\tablewidth{0pt}
\tablecaption{Astrometric Results
\label{tab:ast}}
\tablehead{
  \colhead{}  &
  \colhead{}  &
  \colhead{R.A.\tablenotemark{a}}  &
  \colhead{Decl.\tablenotemark{a}}  &
  \colhead{}  &
  \colhead{}  &
  \colhead{}  &
  \colhead{}  &
  \colhead{}  &
  \colhead{}  &
  \colhead{$\Delta$T} & 
  \colhead{}  &
  \multicolumn{2}{c}{$\pi_{\rm rel}$} &
  \multicolumn{2}{c}{$\mu_{\rm rel}$} &
  \multicolumn{2}{c}{P.A.} &
  \multicolumn{2}{c}{$\pi_{\rm abs}$} &
  \multicolumn{2}{c}{$V_{\rm tan}$} &
  \colhead{} \\
  \colhead{2MASS J} & 
  \colhead{Alternate Name} & 
  \multicolumn{2}{c}{(J2000.0)} &
  \colhead{Cam.\tablenotemark{b}} & 
  \colhead{Filt.} & 
  \colhead{Nf} & 
  \colhead{Nn} & 
  \colhead{Ns} & 
  \colhead{Coverage}  &
  \colhead{(yr)} &
  \colhead{Type\tablenotemark{c}} &
  \multicolumn{2}{c}{(mas)} &
  \multicolumn{2}{c}{(mas yr$^{\rm -1}$)} &
  \multicolumn{2}{c}{(deg)} &
  \multicolumn{2}{c}{(mas)} &
  \multicolumn{2}{c}{(km s$^{\rm -1}$)} &
  \colhead{Notes\tablenotemark{d}} \\
  \colhead{(1)} &
  \colhead{(2)} &
  \colhead{(3)} &
  \colhead{(4)} &
  \colhead{(5)} &
  \colhead{(6)} &
  \colhead{(7)} &
  \colhead{(8)}  &
  \colhead{(9)}  &
  \colhead{(10)}  &
  \colhead{(11)}  &
  \colhead{(12)}  &
  \multicolumn{2}{c}{(13)} &
  \multicolumn{2}{c}{(14)} &
  \multicolumn{2}{c}{(15)} &
  \multicolumn{2}{c}{(16)} &
  \multicolumn{2}{c}{(17)} &
  \colhead{(18)}}

\startdata
 00110078$+$0420245  &  LHS 1032                &  00 11 00.78  &  $+$04 20 25.0  &  1 & ST$-$R  &  106  &   88  &   4  & 1985.71$-$1995.81 &  10.10  & n &   11.72  &  1.25  &   542.1  &  0.5  &  165.0  &  0.2  &   12.37  &  1.25  &  207.6  &   21.0  &      \\
 00192745$+$0450297  &  LHS 1058                &  00 19 27.48  &  $+$04 50 29.9  &  1 & ST$-$R  &  127  &   63  &   3  & 1987.65$-$1995.81 &   8.15  & n &   37.53  &  0.82  &   585.9  &  0.4  &  240.9  &  0.2  &   40.10  &  0.86  &   69.2  &    1.5  &      \\
 00274197$+$0503417  &  PC 0025$+$0047          &  00 27 41.97  &  $+$05 03 41.8  &  2 &  I$-$2  &   98  &   98  &  10  & 1992.75$-$2002.83 &  10.08  & u &   12.42  &  1.56  &    10.3  &  0.3  &   94.8  &  1.5  &   13.24  &  1.56  &    3.7  &    0.4  & 1     \\
 00312326$+$0936169  &  LHS 1089                &  00 31 23.24  &  $+$09 36 17.0  &  1 & ST$-$R  &  125  &   84  &   4  & 1985.62$-$1995.82 &  10.19  & u &    5.44  &  0.84  &   699.4  &  0.4  &  101.0  &  0.2  &    7.19  &  0.93  &  460.9  &   59.6  & 2     \\
 00350768$+$7627544  &  LHS 1103                &  00 35 07.69  &  $+$76 27 54.5  &  1 & ST$-$R  &   91  &   51  &   4  & 1988.93$-$1995.82 &   6.89  & n &   47.69  &  0.52  &   498.5  &  0.2  &   86.0  &  0.2  &   49.22  &  0.55  &   48.0  &    0.5  &      \\
 00361617$+$1821104  &  LSPM J0036$+$1821       &  00 36 16.12  &  $+$18 21 10.4  &  2 &  I$-$2  &  202  &  181  &   9  & 1998.78$-$2007.99 &   9.21  & u &  113.98  &  0.40  &   910.5  &  0.1  &   82.5  &  0.1  &  114.86  &  0.41  &   37.6  &    0.1  & 1     \\
 00470038$+$6803543  &  \nodata                 &  00 47 00.39  &  $+$68 03 54.3  &  3 &  Z$-$2  &   27  &   26  &  50  & 2011.65$-$2015.64 &   3.15  & n &   80.12  &  1.91  &   435.6  &  1.0  &  116.8  &  0.2  &   81.49  &  1.91  &   25.3  &    0.6  &      \\
 00475502$+$4744342  &  LHS 1151                &  00 47 55.08  &  $+$47 44 33.7  &  2 &  I$-$2  &  104  &   87  &  36  & 2008.61$-$2012.89 &   4.28  & n &   27.50  &  0.45  &   859.3  &  0.3  &  126.7  &  0.1  &   28.34  &  0.45  &  143.7  &    2.3  &      \\
 00510351$-$1411047  &  LHS 1157                &  00 51 03.55  &  $-$14 11 05.9  &  3 &  I$-$2  &   91  &   86  &  11  & 2008.64$-$2015.63 &   6.25  & n &   15.69  &  0.50  &   810.6  &  0.2  &  154.7  &  0.1  &   16.65  &  0.51  &  230.7  &    7.1  &      \\
 00554418$+$2506235  &  LHS 1166                &  00 55 44.27  &  $+$25 06 23.6  &  1 & ST$-$R  &  115  &  107  &   5  & 1985.63$-$1995.82 &  10.19  & u &   13.79  &  0.55  &   589.0  &  0.5  &   88.2  &  0.2  &   14.57  &  0.56  &  191.5  &    7.4  & 2     \\
 01001954$+$3224389  &  LHS 1173                &  01 00 19.55  &  $+$32 24 39.0  &  2 &  I$-$2  &   70  &   65  &  23  & 2008.61$-$2012.88 &   4.27  & n &   17.16  &  0.52  &   824.1  &  0.3  &  113.9  &  0.1  &   17.97  &  0.52  &  217.3  &    6.3  &      \\
 01002474$+$1711272  &  LHS 1174                &  01 00 24.70  &  $+$17 11 27.7  &  1 & ST$-$R  &  155  &   73  &   4  & 1983.88$-$1995.82 &  11.94  & u &   12.20  &  0.56  &   910.1  &  0.3  &  124.0  &  0.2  &   13.03  &  0.57  &  330.9  &   14.5  & 2     \\
\enddata
\tablecomments{Table~\ref{tab:ast} is published in its entirety in machine-readable format.  A portion is shown here for guidance regarding its form and content.}
\tablenotetext{a}{Coordinates are from the 2MASS catalog when available, else
  from {\it Gaia} Data Release 1, propagated to epoch J2000 using proper motions
  presented here.}
\tablenotetext{b}{Camera: (1) TI800, (2) TEK2K, (3) EEV24}
\added{\tablenotetext{c}{Type of astrometry: (n) New results (previously unpublished). (u) Update of previously published results, using additional observations. (r) Revision of previously published results using parallax factors with respect to the solar system barycenter and revised corrections from relative to absolute parallax, as described in Section~\ref{sec:evolution}.}}
\tablenotetext{d}{Notes on individual objects: (1) Previous astrometric results were published in
  \citet{2002AJ....124.1170D}.  Results presented here arise from
  updated reduction methods and may include additional data, thus,
  they supercede the results of \citet{2002AJ....124.1170D}. (2) Previous
  astrometric results were published in
  \citet{1992AJ....103..638M}.  Results presented here arise from
  updated reduction methods and may include additional data, thus,
  they supercede the results of \citet{1992AJ....103..638M}.
  (3) Object is a field reference star for LHS 3259
  for which there were measurable astrometric results.  It was named
  LHS 3259.1 in \citet{1992AJ....103..638M}.
  (4) Object is a field reference star for LHS 3548
  for which there were measurable astrometric results.  It was named
  LHS 3548.1 in \citet{1992AJ....103..638M}.}
\end{splitdeluxetable*}

Since the majority of the stars in Table~\ref{tab:ast} were originally identified
due to their large proper motions found in the surveys carried out at
the University of Minnesota \citep{luyten1964} and Lowell Observatory
\citep{1971lpms.book.....G,1978LowOB...8...89G}, researchers in fields studying nearby stars are
often familiar with them by their survey designations.  Column (2) of Table~\ref{tab:ast} gives a common name for the parallax star, and since
197 of them are included in Luyten's LHS Catalog \citep{1979lcse.book.....L},
that designation is given preference.  For 16 proper motion survey
stars with annual motions less than 0.4\arcsec\ Luyten numbers (LP or NLTT)
or Lowell numbers (G) are used.  The names for the remaining stars are
explained in the Notes to Table~\ref{tab:ast}.   For stars for which an earlier
(preliminary) NOFS parallax was presented in either \citet{1992AJ....103..638M} or \citet{2002AJ....124.1170D}, a table note with reference can be found in Table~\ref{tab:ast}.  We emphasize that the determinations
presented here are {\it not} independent of the previous NOFS results but
rather supersede them.

\begin{figure}
\plotone{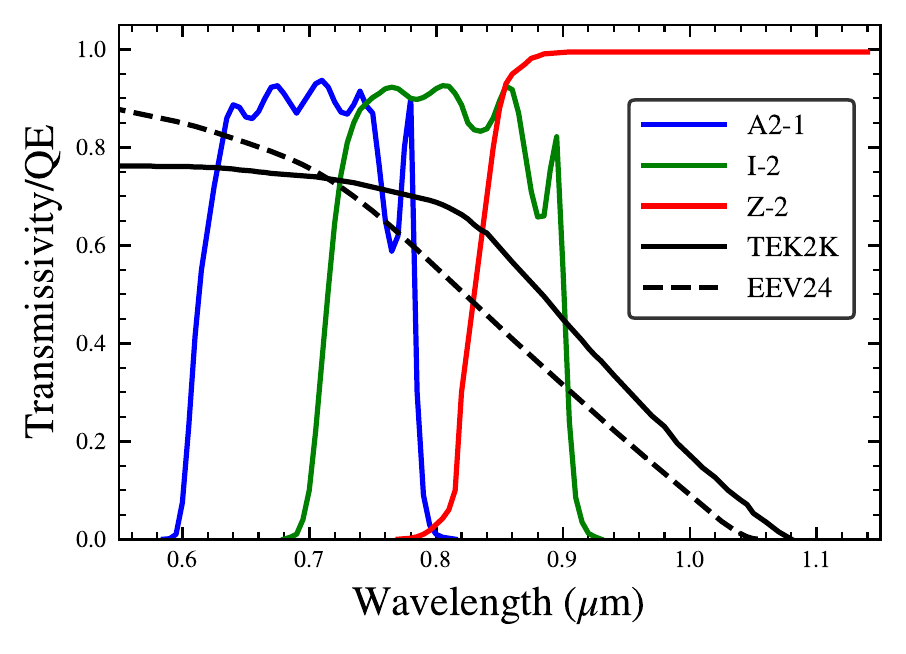}
\caption{Transmissivity of the filters, and quantum efficiency of the CCDs, used in the NOFS parallax program.}
\label{fig-transmission-curves}
\end{figure}

Column (5) indicates the camera employed for each parallax
determination.  Values of 1,2, and 3 correspond to the TI800, TEK2K and EEV24
cameras, respectively.  Column (6) indicates the filter used.  ST-R is the STWIDER filter whose
properties were described in \citet[][Section 2.2]{1992AJ....103..638M}.  A2-1 is an optically flat interference filter that defines a bandpass
centered near 698 nm and with an FWHM of approximately 172 nm.   I-2 is another
optically flat interference filter that defines a bandpass centered
near 810 nm and with an FWHM of approximately 191 nm.  Z-2 is an optically flat 3 mm thick piece of Schott RG830
glass producing a relatively sharp blue-edge cutoff near 830 nm.  The
actual bandpass depends heavily on the sensitivity of the CCD.  The
 bandpass closely approximates the SDSS $z'$ bandpass \citep[cf.][]{1996AJ....111.1748F,2002AJ....123.2121S}.  Figure~\ref{fig-transmission-curves} shows the filter
transmissivities measured by the manufacturer along with the quantum
efficiencies of the TEK2K and EEV24 CCDs supplied by the manufacturers.
The difference between the TEK2K and EEV24 responses is noticeable
and, hence, we did not mix observations taken with different cameras
nor did we mix frames taken with different filters when performing the astrometric reductions.  These filters are
mounted in a tray that is part of the autoguider assembly, placing
them approximately 7 cm in front of the CCDs and assuring that the
out-of-focus image of each star will be insensitive to any slight
filter irregularities. 

Columns (7), (8), and (9) in Table~\ref{tab:ast} give the number of acceptable CCD
frames (observations), the number of separate nights on which those
observations were obtained, and the number of reference stars employed in
astrometric solutions, respectively.  Real-time reduction to centroids is performed
while the observing is in progress, immediately after the camera
shutter closes.  Processing is done for a large selection of potential
reference stars which were designated when the field was originally
added to the program.  Once photometry has been obtained for the field
(parallax target star {\it and} all reference stars), then we carefully
decide which of the potential reference stars will actually be
employed.  Consideration of the frame configuration and reliability of
determining the correction from relative to absolute parallax (see
below) are paramount.  Columns (10) and (11) give the years observed and total epoch range, respectively.  \added{Column (12) specifies whether this is a new result (previously unpublished), an update of a previously published result using additional data, or a revision of a previously published result using parallax factors relative to the solar system barycenter and a revised correction from relative to absolute parallax, as described in Section~\ref{sec:evolution}.}

The derived relative parallax and its \replaced{formal mean (standard) error}{mean (standard) uncertainty} are
given in Column (13) (throughout this work we refer to the trigonometric parallax angle as $\pi$).  The relative
total proper motion and its \replaced{formal error}{uncertainty} follow in Column (14), and the position angle of the proper motion is given in
Column (15).  Our formal solution for the position angle is often precise
to 0.02 -- 0.04 degrees.  However, the reduction of each field's
ensemble of CCD frames to a standard frame does not include the
uncertainty in the orientation of each CCD in its dewar plus an
additional small uncertainty introduced when the dewar is taken on/off
the telescope during monthly instrument changes.  The orientation of
the chip columns (or in the case of the TI800, rows) to rigorously
align with east-west on the sky was carried out by taking star trails
with the telescope pointed at the celestial equator while stationary
on the meridian.  This time-consuming, iterative procedure was carried
out when the cameras were first commissioned.  From then on, scribe
marks on the dewar-to-mounting plate interface were employed for
reorientation following instrument changes.  Variations in solution
plate (frame) constants indicate that quoting the uncertainty in
position angle (Column (15) of Table~\ref{tab:ast}) to only $\pm$0.1 degree
safely accounts for any systematic chip orientation issues.

\begin{figure}
\plotone{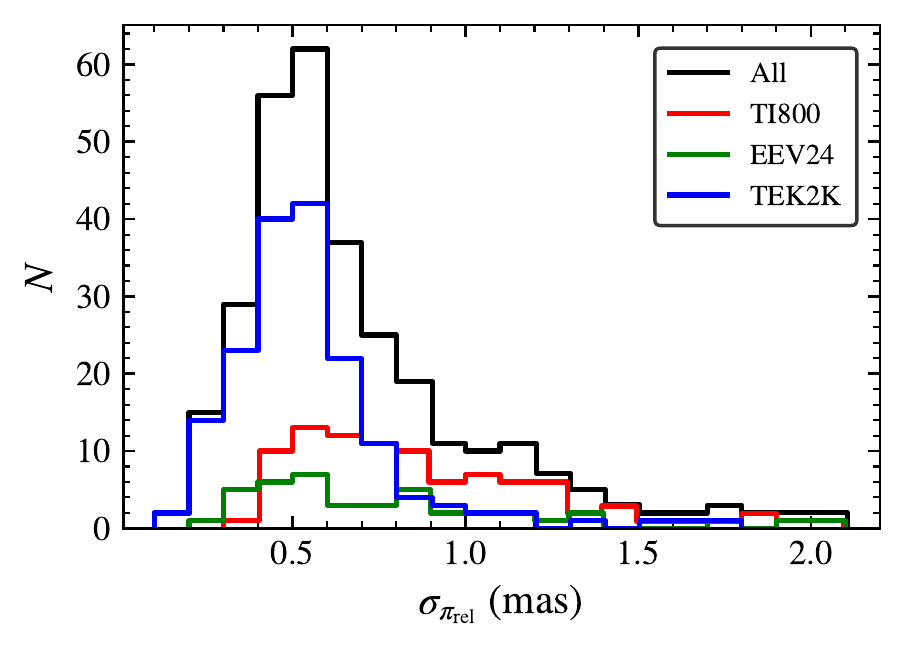}
\caption{Distribution of the relative parallax \replaced{errors}{uncertainties}, for all stars in this paper (black histogram), and
for stars observed with the individual cameras.}
\label{fig:rel-pi-errors}
\end{figure}

Figure~\ref{fig:rel-pi-errors} shows the frequency distribution for the \replaced{formal errors}{uncertainties} of the
relative parallax determinations, including a breakdown for the three
separate cameras.  All three cameras produced relative parallaxes with formal precisions in
the $\pm0.3 - 0.7~{\rm mas}$ range when $\gtrapprox30$ observations spanning an epoch range of $\gtrapprox4$ years were employed.  The extended tails toward larger \replaced{error}{uncertainty} in the Figure~\ref{fig:rel-pi-errors} distributions for the TEK2K and EEV24 cameras primarily result from very faint targets or very bright targets where the exposure times are too short to properly average out atmospheric effects.  The more pronounced \replaced{exess}{excess} of larger \replaced{errors}{uncertainties} for the TI800 fields is primarily due to very poor reference star frame configurations available in that camera's limited FOV.

The derived absolute parallax and its \replaced{formal mean error}{uncertainty} are presented
in Column (16).  The correction
from relative to absolute parallax was carried out using the
procedures described in detail by \citet{2016AJ....152..118H}.  In summary, a
photometric parallax is derived for every reference star employed
using a calibrated $M_V$ vs $V-I$ diagram constructed from large, accurate
{\it Hipparcos} parallaxes for stars with $M_V<13$, plus stars with $M_V>9$ based
on accurate USNO photographically measured parallaxes.  Photometry
measured at NOFS on the Johnson-Kron-Cousins system is employed (see Section~\ref{section:photometry}).
For fields located at lower galactic latitudes, other reference star
photometry (2MASS $JHK_s$; $BVI$ from various sources) is employed to help
resolve dwarf/giant ambiguities and improve estimates for interstellar
reddening/extinction corrections.

\begin{figure}
\plotone{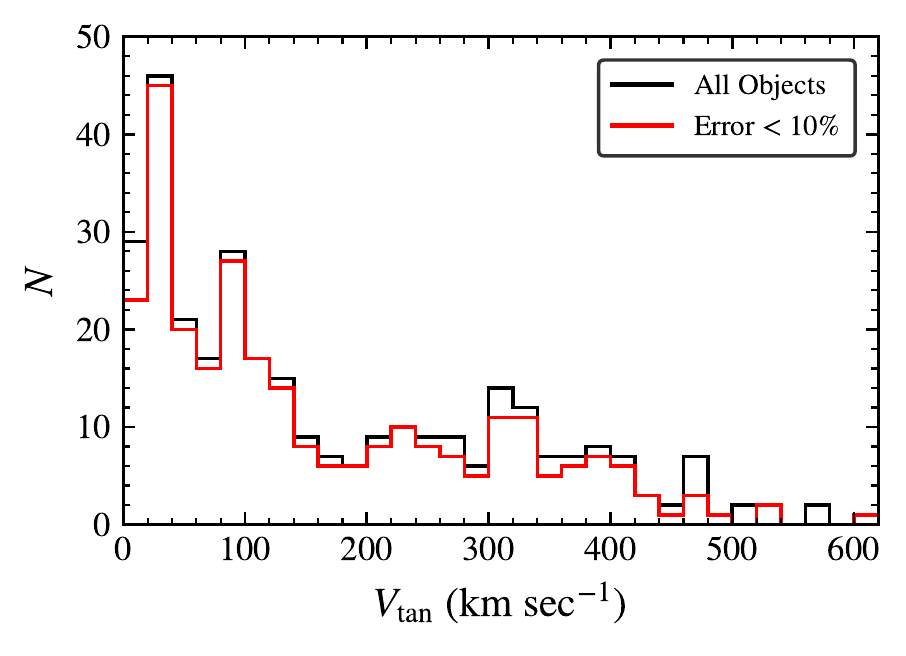}
\caption{Distribution of tangential velocities.}
\label{fig-vtan}
\end{figure}

The calculated tangential velocities and their \replaced{errors}{uncertainties} are given
in Column (17) of Table~\ref{tab:ast}.  We
have not attempted to convert from relative to absolute proper
motions.  Such corrections are very uncertain, due to necessary
assumptions regarding galactic rotation (see \citealt{2016AJ....152..118H}, Section 3.3).  However, for the faint reference stars employed
for the parallax determinations presented here, these corrections
amount only to 1 -- 3 mas~yr$^{-1}$. Since all but 19 of the stars in the
present sample have relative proper motions greater than 100 mas~yr$^{-1}$,
the use of $\mu_{\rm rel}$ in calculating $\vtan$ has little effect.  Figure~\ref{fig-vtan}
shows the frequency distribution of $\vtan$ for the Table~\ref{tab:ast} sample.  Note
that fully one-third of the stars in the Table~\ref{tab:ast} sample have $\vtan>200~
{\rm km~s^{-1}}$.  This is, of course, not surprising since our sample is highly
biased toward higher tangential velocities since targets were given
priority to being added to our parallax program if we knew or
suspected that they might be subdwarfs.

\subsection{Individual Stars}
Many of the stars in Table~\ref{tab:ast} are among the most intensely studied objects in the nearby solar neighborhood.  It is beyond the scope of the present paper to review the extensive literature for them.  The interested reader is referred to relevant references in SIMBAD.  However, some brief notes concerning our Table~\ref{tab:ast} astrometry for a few stars follows.

\label{sec:stars}
{\bf PC 0025+0447} and\\
{\bf 2MASS J16241436+0029158} were both observed with the I-2 filter
before the Z-2 filter became available.  The large \replaced{mean errors}{uncertainties} of the
relative parallaxes for these two stars reflects the marginal exposure level of these very red and faint targets.

{\bf 2MASS J00470038+6803543} has a large relative parallax \replaced{error}{uncertainty}, the direct
result of the faintness of this object even at $Z$-band.  Observations
could only be made in subarcsec seeing and on nights with clear,
dark sky.

{\bf LHS 1252} has a very poor available reference star frame,
necessitating the use of one very faint star and a faint galaxy.

{\bf GAT 1370} has a preliminary distance estimate -- $d = 2.4(+0.7,-0.40)$~pc --
derived by \citet{2003ApJ...589L..51T} and the star is often
referred to as ``Teegarden's star.''   The weighted mean of five independent parallax determinations (see Table~\ref{tab:astcomp} below) places this star at a distance of $3.842 \pm 0.004$~pc, making it the 23rd nearest non-BD star (system) from the sun.

{\bf 2MASS J04193697+1433329},\\ {\bf 2MASS J04223075+1526310},
and\\ {\bf 2MASS J04325119+1730092} were suggested as possible
brown dwarf members of the Hyades cluster \citep{1989MNRAS.238..145L} but
were shown by NOFS astrometry \citep{1999AJ....117..339H} and spectroscopy
\citep{1999AJ....117..343R} to be pre-main-sequence stars in the background
of the cluster.

{\bf 2MASS J04433761+0002051} was suggested as an ultracool late
M-dwarf member of the AB Doradus moving group by \citet{2012AJ....143...80S}.  Earlier, \citet{2009AJ....137....1F} had proposed it as a
member of the $\beta$ Pictoris group and \citet{2014ApJ...783..121G}
concurred.  The parallax determination by \citet{2016ApJ...833...96L} supports
the $\beta$ Pictoris membership.  Our Table~\ref{tab:ast} results are in good
agreement (see Table~\ref{tab:astcomp} below) and adds further strength to that
interpretation.

{\bf 2MASS J04435686+3723033}, the BY Draconis variable V961 Per, was
identified as a likely member of the $\beta$ Pictoris moving group
by \citet{2010AJ....140..119S} who gave an estimated kinematic distance
of 76.9 pc.  Our Table~\ref{tab:ast} parallax translates to a distance of 73.7
(+2.6, -2.3) pc, in satisfactory agreement with their prediction.
\citet{2010AJ....140..119S} also report a faint common proper motion
companion located $\sim 9\arcsec$ to the east of 2M0443+3723.  We performed a
solution using 23 uncontaminated exposures of the companion, obtaining
$\piabs = 14.41\pm1.31$~mas and $\murel = 64.5~\masyear$ at $\pa=158.0 \pm 0.5{^\circ}$.  Both strongly support the physical nature of the pair.
At epoch 2015.053 we measure $\rho =7.68\arcsec$ at $\pa = 91.65^{\circ}$.

\begin{figure*}
\plotone{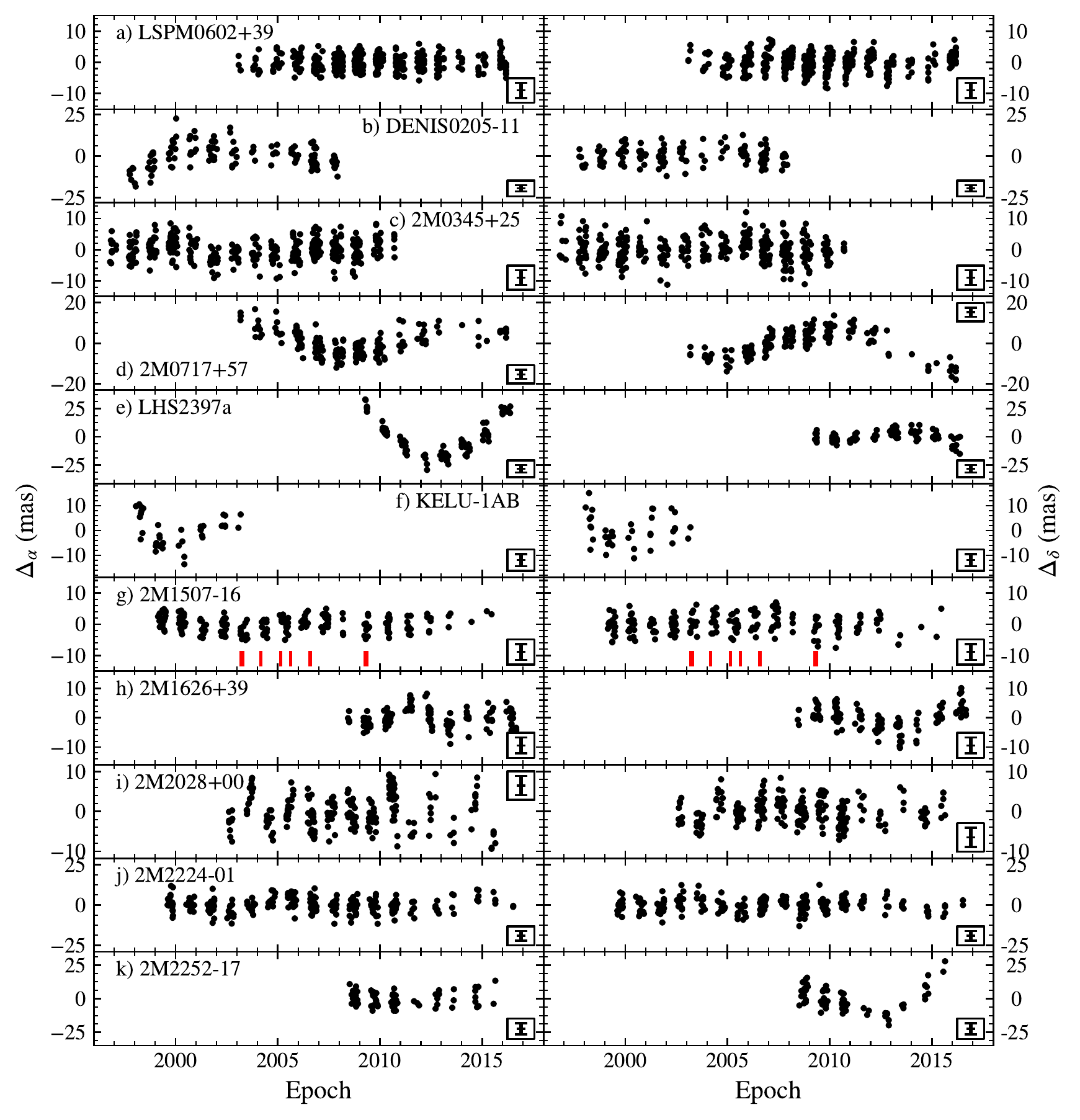}
\caption{Astrometric residuals for individual stars discussed in the text.  The left panels show the residuals in right ascension, while the right panels show the residuals in declination.  Typical error bars are indicated on the right-hand side of each panel.  For 2MASS J15074769-1627386 (Panel g), the red lines indicate the epochs of the radial velocity measures of \citet{2010ApJ...723..684B}.}
\label{fig:residuals}
\end{figure*}

{\bf LSPM J0602+3910}, an L1V star, was maintained on the observing program for
over 12 years to monitor the stability of our astrometric results and
to look for any detectable astrometric perturbation.  The
relative parallax \replaced{error}{uncertainty} of 0.19~mas is the third best achieved by
the CCD program.  The formal residuals from the solution in each
coordinate (Figure~\ref{fig:residuals}, Panel~a) show no convincing evidence of an astrometric perturbation and demonstrate the stability of residuals derived from observations spanning epoch ranges as long as a decade.

{\bf LHS 1839} has a proper motion which has carried it toward a brighter field
star.  During the epochs of our astrometric measures (1986.23 to
1995.16) they were still well enough separated to yield uncontaminated
measures when the seeing was $\sim1.5$\arcsec\ FWHM or better.

{\bf LP540--16} has a proper motion $>1.0\arcsec~{\rm yr}^{-1}$,
first announced by \citet{1983PMMin..63....1L}, and was added to the NOFS TEK2K
program in 2003 March.  It was removed as completed in 2008 March.
\citet{2002AJ....124.1190L} announced (what turned out to be) a
``rediscovery'' of the same star, and it was inadvertently added to
our EEV24 program in 2012 December.  When the repeat was discovered in
2014 March, it was again deactivated.  The excellent agreement between
the two totally independent determinations is reassuring.  Thus, a weighted mean of $14.55 \pm 0.23$~mas is appropriate.

{\bf LHS 2471} has a very large \replaced{error}{uncertainty} in the parallax determination,
primarily due to the extremely poor reference frame available.

{\bf LHS 2557} was observed over two intervals -- from 1993.07 to 1997.29 and
then again from 2001.28 to 2008.17 -- due to the star's proper motion
carrying it across a faint field star.

{\bf LHS 453} has a revised solution in Table~\ref{tab:ast} employing a tighter
4 star reference frame and over twice the number of observations as
were available for the preliminary parallax determination presented in
\citet{1992AJ....103..638M}.

{\bf LSPM J1826+3014} has a proper motion that carried this star across a field
galaxy making the target star unobservable between 2005.37 and
2008.31.

{\bf LHS 3406}, observed with TI800 where a rich but large reference star
frame was available, was also observed using TEK2K as a
consistency check.  As Table~\ref{tab:ast} shows, the two independent absolute
parallax determinations agree very well with each other to within the
combined \replaced{formal errors}{uncertainties}.  Thus, a weighted mean of $69.09 \pm 0.38$~mas is appropriate.

{\bf LHS 474} is vB10 and our preliminary parallax determination was
presented in \citet{1992AJ....103..638M}.  Continued observations were
complicated by vB10's proper motion carrying it across a moderately
bright field star, and it was removed from the program on 2009 August 21.  The
residuals from that parallax solution showed no sign of a perturbation
in either RA or DEC.  When the astrometric discovery of a purported
candidate planet was announced \citep{2009ApJ...700..623P}, vB10 was added
to the EEV24 program. Subsequently, \citet{2010ApJ...711L..19B} presented
precision ($\pm$~10 m~s$^{-1}$) radial velocity measures spanning $\sim225$ days
in 2009 and found no variation, concluding that the proposed 6.4
Jupiter-mass planet in a $P=0.744~{\rm yr}$ orbit is ``spurious.''  Our EEV24
observations now span the 7.06 yr interval from 2009 June 15 to 2016 July 5.
However, vB10 passed very close to another field star in the 2010 -- 2011 interval.  Nonetheless, these observations show no indication of the Pravdo \&
Shaklan perturbation.
However, as Pravdo \& Shaklan
themselves noted, this field is ``an astrometrist's dream in a
nightmarish setting.''  On the positive side, it's location at $b=-3.3^{\circ}$
provides a plethora of potential reference frame stars.  On the
negative side, at least for the parallax determination, is that its
location near the galactic equator presents problems with interstellar
reddening/extinction in determining the correction from relative to
absolute parallax.  Furthermore, vB10's large proper motion through
such a crowded field presents potential for contamination at repeated
epochs.  At this time, we cannot reject such possibilities and
are continuing our observations.

{\bf LHS 3548} and\\ {\bf 2MASS J20253876$-$0612094} have updated TI800 astrometry employing a very tight 5 star reference frame.  Note that the latter was referred to as LHS 3548.1 in \citet{1992AJ....103..638M} because it was found to have a measurable parallax serendipitously while serving as a reference star.

{\bf LSPM J2036+5100} has an incorrectly reported position angle for the proper motion of
100.8$^{\circ}$ in \citet{2003AJ....125.1598L}.

{\bf LHS 3684} uses only the RA parallax solution in Table~\ref{tab:ast}, due to some contamination from a faint star $\sim2$\arcsec\ due
south of the target.  Additionally, we suspect the quoted proper motion might also
be affected.

{\bf 2MASS J21374019+0137137} was suggested as a likely member of the
$\beta$ Pictoris moving group by \citet{2012AJ....143...80S}.  Treating
apparent radial velocity variations as due to an SB2 led to an
unphysical interpretation.  In our good seeing observations ($\le0.8\arcsec$
FWHM) the images of 2M2137+0137 seem to be slightly distorted when
compared to nearby field stars.  This suggests that it might be a
close binary where the faint companion has a separation of $\sim 0.5\arcsec$.

{\bf 2MASS J22004158+2715135} was observed at the request of J. Schlieder 
for his studies of moving group membership candidates.  Because of the
brightness of the target ($I = 9.59$), exposure times were restricted to
$< 20$~s in good seeing.  Consequently, the averaging over atmospheric
effects was inadequate, and the resulting \replaced{errors}{uncertainties} in the solution reported
in Table~\ref{tab:ast} are large.  In seeing $\le 0.8\arcsec$ FWHM a faint star is detected
roughly due north of 2M2200+2715.  At epoch 2012.712 we measure it at
$\rho  =5.34\arcsec$ and $\pa = 357.2^{\circ}$.  Common proper motion has not yet been
established for the two stars.

{\bf LHS 541} is vB12, a companion to LHS 540 = BD-14:6437AB at
$\rho \sim 15\arcsec$ and $\pa \sim 47^{\circ}$.  The Table~\ref{tab:ast} solution is restricted to
observations taken in $\le 1.6\arcsec$ FWHM to preclude contamination from
LHS 540.

{\bf LHS 3937} was found by \citet{1980nltt.bookQ....L} to have $\mu=803~{\rm mas~yr^{-1}}$ at $\pa =104^{\circ}$,
whereas \citet{2003ApJ...582.1011S} reported $\mu=990~{\rm mas~yr^{-1}}$ at
$\pa =106.8^{\circ}$.
Our Table~\ref{tab:ast} result ($\mu=766.4\pm0.3~{\rm mas~yr^{-1}}$ at
$\pa =104.9\pm0.1^{\circ}$) is in closer agreement with that of Luyten.

\subsection{Resolved Binary Systems -- Physical and Optical}

Several of the stars in Table~\ref{tab:ast} are members of binary systems whose
components are separated enough to be resolved in our observations.
Comments on them follow.

{\bf LHS 1203} and {\bf LHS 1205} form a common proper motion pair with separation $\rho \sim 7.9\arcsec$ at
$\pa \sim 133.6^{\circ}$.  \citet{2002ApJS..141..187B} reported significantly different
proper motions for the two.  Our astrometry confirms the physical
nature of the pair.

{\bf PG 0235+064B} was discovered as a common proper motion companion to the DA3.5
white dwarf PG0235+064 = GJ3173 = 2MASS J02383078+0638071 in the
course of our astrometric observations of the WD.  For the WD we
measure $\pirel=16.03\pm0.33~{\rm mas}$, $\mu_{\rm rel} = 86.3~{\rm mas~yr^{-1}}$ at $\pa =191.2^{\circ}$ which support the physical nature of the pair.

{\bf LHS 189} and {\bf LHS 190} form a close physical binary ($\rho \sim 1.2\arcsec$, $\pa \sim 146.2^{\circ}$) which required DAOPHOT centroiding of the targeted stars to avoid
contamination between the two.  The fainter of the two (LHS 190) could
not be measured on 34 of the frames suitable for the brighter
component (LHS 189).

{\bf LHS 2099} and {\bf LHS 2100} form a common proper motion pair with $\rho = 6.66\arcsec$ and
$\pa = 100.8^{\circ}$.  LHS 2100 is approximately 2.8 mag fainter than
LHS 2099 in the STWIDER-band used in our astrometry, resulting in
consistently weakly exposed images.  This accounts for the unusually
large \replaced{error}{uncertainty} in our parallax determination.  Nevertheless, our
astrometry is good enough to confirm the physical status of the
system.

{\bf LHS 2140} is the bright ($V \sim15.04$) primary component of a physical binary
system with LHS 2139 ($V \sim19.6$; spectral type DC? from \citealt{1997PASP..109..849G}).  From observations obtained in good
seeing we measure $\rho = 6.7\arcsec$, $\pa = 28.9^{\circ}$.  From 100 frames suitable
for measurement of the fainter component we find $\pirel=13.73\pm1.40~\mas$
and $\murel=580.0\pm0.7~\masyear$ at $\pa = 186.9\pm0.7^{\circ}$.  The
system is clearly physical.

{\bf LHS 6176} is the bright component in a common proper motion binary system containing a
T8p secondary star $\sim52$\arcsec\ away \citep{2012ApJ...760..152L, 2013MNRAS.433..457B}.  The secondary is much too faint in the optical for us to observe.

{\bf LHS 2444} and {\bf LHS 2445} form a common proper motion pair with $\rho = 10.74\arcsec$ and
$\pa = 137.6^{\circ}$.  Our astrometry confirms the physical status of
the system.

{\bf G166-37} (= Ross 50 = LP381-87) and {\bf LP 381-86} form a widely separated
common proper motion pair.  We measure $\rho = 200.8\arcsec$ at $\pa = 205.8^{\circ}$
which, along with our parallaxes, confirms the physical status of the
system at the 1.5-sigma level.

{\bf LHS 3001} and {\bf LHS 3002} are clearly a physical system.  Due to the
faintness of the secondary (LHS 3002) astrometry could not be obtained
from 11 of the frames employed for LHS 3001.

{\bf LSPM J1457+2341S}\\ (=2MASS J14572616+2341227), located $\sim$~2.3\arcsec\ east and
$\sim$~3.2\arcsec\ south of the dwarf carbon star candidate SDSS 1457+2341 (=2MASS
J14572597+2341257) \citep{2013ApJ...765...12G} is a possible -- but far from certain
-- physical companion.  Being $\sim2.3$ mag fainter than the dC star at
$I$-band and only $\sim4$\arcsec\ distant, it is difficult to measure.  For the
brighter component we measure $\pirel=7.42\pm0.55~\mas$ and
$\murel=359.1\pm0.3~\masyear$ at $\pa = 260.0\pm0.1^{\circ}$ from 35
observations.  Although not definitive, the close similarity of our
values is strongly suggestive.  Recently, \citet{2016ApJS..224...36K}
announced this as a candidate physical triple system based on data
from the ALLWISE motion survey.  Our observations are continuing.

{\bf LHS 3181} appeared to be a binary on several exposures taken in very good seeing.  The potential companion is faint ($\Delta_r = 3.73 \pm 0.03\ {\rm mag}$ in the ST-R bandpass) and located at $\rho \sim 3.56\arcsec$ and $\pa \sim 155.4^{\circ}$ (epoch 1991.44).  Observations continuing until 1995.57 confirmed that this faint star is a background field star.

{\bf 2MASS J17054834-0516462} appeared to have a binary companion with $\rho \sim 1.36\arcsec$ and
$\pa \sim 5^{\circ}$ in NICMOS exposures on {\it HST} \citep{2006AJ....132..891R}.  However,
\citet{2011AJ....141...54A} concluded from photometry that the faint star was
most probably a more distant ($\sim$~200~pc) late M dwarf.  Our results in
Table~\ref{tab:ast} are from DAOPHOT centroids of the 2MASS target.  Attempts to
measure the fainter star were only marginally successful.  Employing
84 acceptable observations made in better seeing, we find a proper
motion that is immeasurably small and an absolute parallax of
$6.3\pm4.9$ mas, a value that confirms it is not a physical companion.

{\bf 2M19302746-1943493} has a faint star nearby for which we estimate $\rho \sim 1.5\arcsec$,
$\pa \sim {\rm SSW}$, and $\Delta_I=0.8~{\rm mag}$.  With the field located at a large
zenith distance, the pair is often significantly blended on our
frames.  DAOPHOT centroiding was employed for both stars to determine
whether they form a physical pair.  For the fainter star we find
$\pirel=39.34\pm1.67~\mas$ and $\murel=216.7\pm0.6~\masyear$ at $\pa =
104.5\pm0.2^{\circ}$.  Allowing for the problem with image blending, it
appears that they do form a physical pair.  Based on our estimated
value of $\Delta_I=0.8$, we infer that the B component is most likely an
$\sim$~M8V companion to the M6.5Ve primary component.

{\bf LHS 5359} and {\bf LHS 5360} are clearly a physical system.  The fainter
component (LHS 5360) is located $\sim2.5\arcsec$ away from the considerably
brighter component and is measurable on only $\sim$~43\% of the frames
acceptable for LHS 5359.

\subsection{Unresolved Binary Systems -- Old and New}
\label{section:unresolved-binaries}

Table~\ref{tab:ast} includes 12 binaries for which we can not resolve the
individual components from our ground-based observations.  Four
are previously unrecognized systems.

{\bf DENIS J0205.4-1159AB} was discovered to be a close physical binary pair by
\citet{1999ApJ...526L..25K}.  Using a collection of data from various
sources, \citet{2003AJ....126.1526B} estimated the period to be $\sim75$ yr.  The
residuals from our astrometry (Figure~\ref{fig:residuals}, Panel~b) cover 10.17 years over the
epoch range from 1997.8 to 2007.9 and show a possible inflection
(nonlinearity) in RA starting around 2001.

{\bf 2MASS J03454316+2540233} was observed to be a double-lined spectroscopic binary by
\citet{1999ApJ...521..613R}.  The residuals from our 13.91 yr of
observational coverage (Figure~\ref{fig:residuals}, Panel~c) show some evidence for a low-amplitude perturbation in the residuals with an estimated period of
$\sim7.7$ yr.  However, the evidence is admittedly weak and observations
are not being continued on this field.

{\bf 2MASS J07171626+5705430} exhibits residuals from our solution for parallax and proper
motion clearly indicating a perturbation in both RA and DEC (Figure~\ref{fig:residuals}, Panel~d).
Since we apparently have not observed a full period yet, all we can
say for sure is that $P>10~{\rm yr}$.

{\bf 2MASS J08503593+1057156} is the star selected by \citet{1999ApJ...519..802K}
to define the L6V spectral class.  Notable was the presence in the spectrum
of strong (${\rm EW} \sim 15$ \AA) Li $\lambda 6708$\AA\ absorption.  {\it HST} images of the star
by \citet{2001AJ....121..489R} revealed that it is a tight binary with $\rho \sim 0.16\arcsec$,
$\pa \sim 250^{\circ}$, and $\Delta_I \sim 1.3~{\rm mag}$.  They also noted the presence of a
slightly brighter field background M dwarf located approximately due east of
2M0850+1057 with $\rho < 2\arcsec$ on 2000 February 1 Feb.  The preliminary NOFS CCD parallax
$\piabs = 39.1\pm 3.5~{\rm mas}$ presented in \citet{2002AJ....124.1170D} was based on 30
observations taken between 1997 December 4 and 2001 March 29.  A preliminary NOFS near-infrared
parallax based on 13 observations with a mean epoch of 2001.79 found
$\piabs = 26.22 \pm 4.21~{\rm mas}$ \citep{2004AJ....127.2948V} and the discrepancy between the
two determinations was noted.  \citet{2011AJ....141...71F} have argued quite
convincingly that this poor agreement is most likely due to contamination
in the earlier CCD observations from the field M dwarf which we measure to
have small relative proper motion such that 2M0850+1057 is moving away from
it at $\sim ~142~\masyear$.  The astrometry presented here in Table~\ref{tab:ast} was obtained by
rejecting all observations taken prior to 2001 February 17 and all observations where
the seeing was greater than $1.4\arcsec$~FWHM.  Unfortunately, this left only 25
acceptable frames and the large \replaced{error}{uncertainty} in the derived relative parallax
reflects the paucity of observations for this very faint target.  However,
our CCD determination is now in satisfactory agreement with both the improved
NOFS near-infrared parallax and with the near-infrared parallax reported by \citet{2012ApJS..201...19D}.  See Table~\ref{tab:astcomp} below.

{\bf 2M09153413+0422045} is a pair of L7V stars with $\rho = 0.73\arcsec$ \citep{2006AJ....132..891R}.  Attempts to measure both stars from frames obtained in
very good seeing were not successful.  Using only frames obtained in
moderate seeing (generally, 1.2 -- 1.8\arcsec\ FWHM), where the pair is well
blended, yielded the results reported in Table~\ref{tab:ast}.

{\bf LHS 2397a} was announced to have an unresolved brown dwarf companion by
\citet{2003ApJ...584..453F}.  Our TI800 parallax for this star was presented
as a completed determination well before this \citep{1992AJ....103..638M}.
The large parallax \replaced{error}{uncertainty} in that result is mostly a consequence of the
very poor reference star frame available, necessitating the use of a
very faint galaxy.  The field was added to our EEV24 program (where it
benefits from a much more favorable reference star frame) in 2009 April.  With $P\sim14.2~{\rm years}$ \citep{2009ApJ...699..168D,2010ApJ...711.1087K},
we have observed it to date for less than half of a full orbital
period.  However, the residuals from our parallax solution shows clear
evidence for an astrometric perturbation, especially the RA coordinate
(Figure~\ref{fig:residuals}, Panel~e).  The large (and growing!) \replaced{error}{uncertainty} for the relative parallax
presented in Table~\ref{tab:ast} reflects our current inability to remove the
perturbation from our EEV24 distance determination.  \citet{2017arXiv170305775D} recently analyzed their measures of the photocentric motion for this system and derived a mass ratio of $0.71 \pm 0.03$ for the components.

{\bf Kelu-1AB} was identified as a physical pair of L dwarfs by \citet{2005ApJ...634..616L} using laser-guided adaptive optics on the Keck II
Telescope.  In the near-infrared they measured $\rho =0.291\pm0.002\arcsec$ and $\pa = 221.2\pm0.6^{\circ}$.  A roughly estimated period of 35 years was obtained, although
a much larger range of possible periods ($\sim$ 15 -- 105 years) was
acknowledged.  Our observations span only 5.12 years and the residuals
from the parallax/proper motion solution are shown in Figure~\ref{fig:residuals}, Panel~f.
There is a suggestion of an astrometric perturbation in
our data.  However, the meridional crossing zenith distance of the
field is over 60$^{\circ}$, the total number of acceptable observations is
small, and the TEK2K camera was employed at $I$-band (resulting in many
weak exposures), so the possibility of a false perturbation can not be
ruled out.  The rather poor agreement between the three currently available parallax determinations for this system can be seen in the Table~\ref{tab:astcomp} compilation below.

{\bf 2MASS J15074769-1627386} was added to our TEK2K program on 1999 February 23 and the
residuals from our parallax solution in RA show the suggestion of a
low-amplitude perturbation (Figure~\ref{fig:residuals}, Panel~g).  Inflections (nonlinearities) in
the years 2003, 2008, and 2012 suggest an orbital period of $\sim$~9 -- 10
years.  Seven high-precision (rms error $\pm0.27~\kmsec$) infrared
radial velocity measures by \citet{2010ApJ...723..684B} showed a statistically
significant linear variation of $\sim0.5~\kmsec$ over the time integral
2003.20 to 2006.58.  Then, three measures between 2009.26 and 2009.36
indicated either a flattening of this variation or possibly a
reversal.  These authors suggest that such behavior -- if real --
would indicate the presence of a low-mass companion in an orbit with
$P>13~{\rm years}$.  The time coincidence of their event with ours motivates
us to continue monitoring this target.  Our observations now cover an
epoch range of 16.3 years and predict that another inflection should
occur in the 2016 -- 2017 time frame.  The large parallax which we
measure for this star should not be significantly affected by a
potential perturbation this small.

{\bf 2MASS J16262034+3925190} is the only sdL observed on the NOFS optical parallax program.
Residuals from the parallax solution show evidence for a low-amplitude
perturbation in both RA and DEC (Figure~\ref{fig:residuals}, Panel~h).  The period appears to be in
the $8\pm1$ year range but since our observations only cover an epoch
range of 8.07 years, that is very uncertain.  Observations are
continuing on this field.

{\bf 2MASS J20282035+0052265} displays evidence for a low-amplitude, short-period ($\lesssim2~{\rm yr}$)
perturbation in both the RA and DEC residuals to the parallax
solution (Figure~\ref{fig:residuals}, Panel~i).  Observations are continuing.  {\it Gaia} will
easily confirm or refute the reality of the perturbation.

{\bf 2MASS J22244381-0158521} was added to our TEK2K program on 1999 February 23 and observations
have continued for 16+ years now.  The residuals from the parallax
solution presented in Table~\ref{tab:ast} (Figure~\ref{fig:residuals}, Panel~j) show evidence for a low-amplitude perturbation in both RA and DEC with an estimated period of
10 -- 12 years.  Our observations are continuing.

{\bf 2MASS J22521073-1730134} was resolved in HST NICMOS images by \citet{2006ApJ...639.1114R}
as an L+T-type binary system with $\rho = 0.130\pm0.002\arcsec$ at
$\pa = -9.6\pm1.2^{\circ}$.  They estimate that the period would lie
somewhere in the range 3 -- 38 years.  Our observations cover an epoch
range of 7.1 yr and show definite evidence for a perturbation in DEC
(Figure~\ref{fig:residuals}, Panel~k).  Observations are continuing.

\subsection{Comparison with Other Parallax Determinations}

\begin{figure}
\plotone{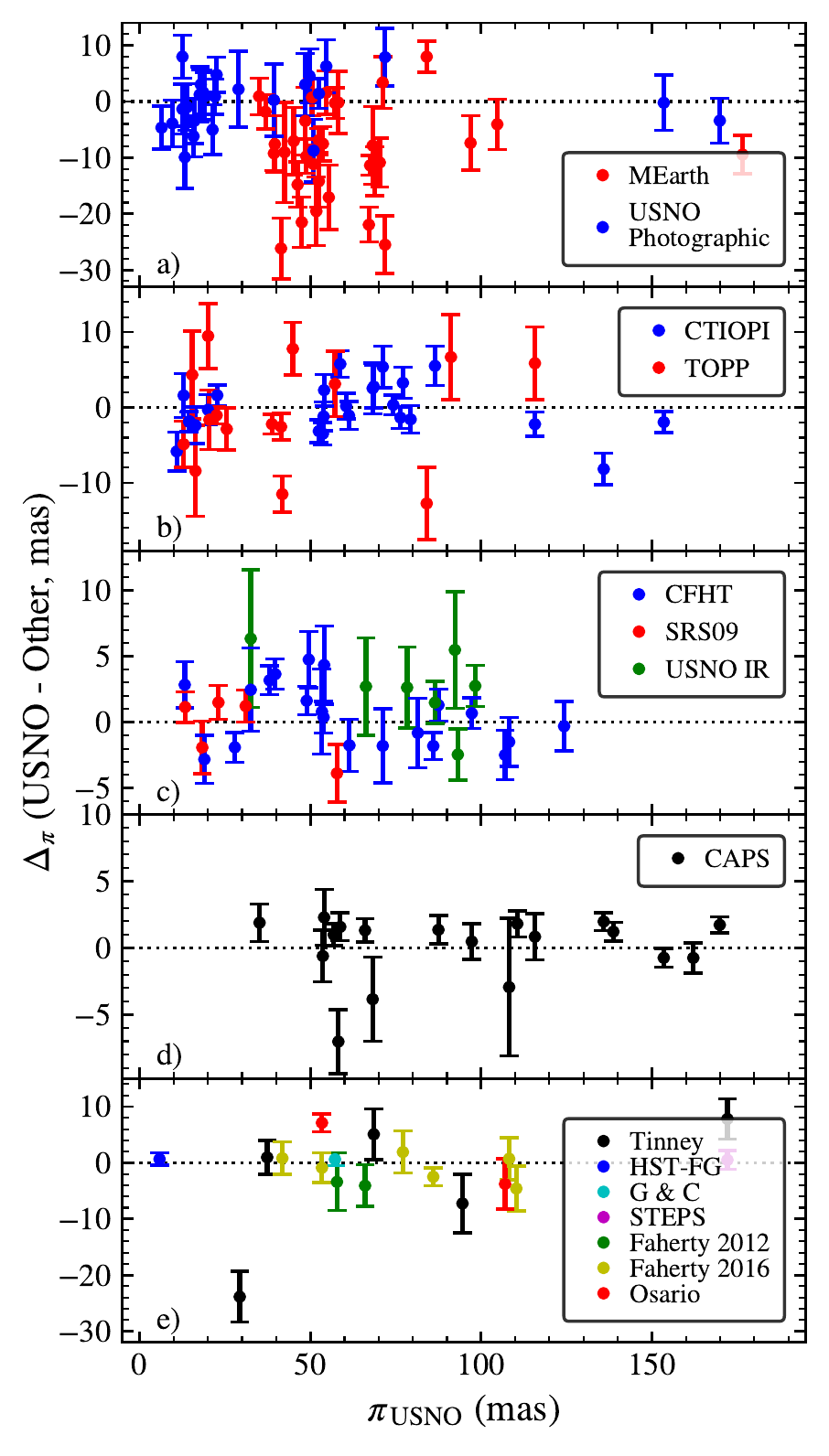}
\caption{Differences in the parallaxes for stars in common between those presented in this paper and various other programs.  Panel (a): MEarth (red) and USNO photographic (blue) parallax programs.  Panel (b): RECONS/CTIOPI (blue) and TOPP (red).  Panel (c): Infrared parallaxes from \citet{2012ApJS..201...19D} and \citet[][blue]{2016ApJ...833...96L}, \citet[][red]{2009A+A...493L..27S}, and the USNO ASTROCAM parallax program (\citealt{2004AJ....127.2948V}, green).  Panel (d): CAPS parallax program \citep{2016AJ....152...24W}.  Panel (e): \citet{1995AJ....110.3014T} and \citet[][black]{1996MNRAS.281..644T}, \citet[][blue]{1997AJ....114.1268M}, \citet[][cyan]{2009AJ....137..402G}, \citet[][maroon]{2009ApJ...700..623P}, \citet[][green]{2012ApJ...752...56F}, \citet[][red]{2014A+A...568A...6Z}, and \citet[][yellow]{2016ApJS..225...10F}.}
\label{fig:compare}
\end{figure}

Trigonometric parallax determinations have already appeared in the
literature for 124 of the stars in Table~\ref{tab:ast}.  Table~\ref{tab:astcomp} presents a
compilation of these results for comparison.  The largest overlap with
our Table~\ref{tab:ast} determinations is with MEarth \citep{2014ApJ...784..156D} where
we have 33 stars in common.  Since the published proper motion
determinations for MEarth did not include estimated errors for each
star individually, we have adopted values of $\pm1.0~\masyear$ for each
coordinate based on advice from Dittmann (2016, private communication).  The second largest
overlap is with USNO photographically determined parallaxes, numbering
25 stars in common.  (A compilation of all USNO photographic values can be
found in Vol.II of YPC4.)  Figure~\ref{fig:compare}, Panel~a shows a comparison of the values
for these two subsets with our CCD values.  The stars with USNO
photographic determinations are some of the faintest targets attempted
at NOFS.  The \replaced{formal mean parallax errors}{parallax uncertainties} are large, with a median
value of $\pm$4.2~mas.  Examination of Figure~\ref{fig:compare} shows that most of the
two independent USNO determinations agree to within 1.5$\sigma$ of the
combined \replaced{formal error}{uncertainty} in the difference, and hence, support the
reliability of USNO photographic parallaxes at the precision level
published.

\begin{deluxetable}{cr@{$\pm$}rr@{$\pm$}rr@{$\pm$}rr}
\tabletypesize{\tiny}
\tablewidth{0pt}
\tablenum{2}
\tablecolumns{2}
\tablecaption{Astrometric Comparisons
\label{tab:astcomp}}
\tablehead{
  \colhead{} &
  \multicolumn{2}{c}{$\pi_{\rm abs}$} &
  \multicolumn{2}{c}{$\mu_{\rm rel}$} &
  \multicolumn{2}{c}{P.A.$_{\rm rel}$} &
  \colhead{} \\
  \colhead{2MASS J} &
  \multicolumn{2}{c}{(mas)} &
  \multicolumn{2}{c}{(mas yr$^{\rm -1}$)} &
  \multicolumn{2}{c}{(deg)} &
  \colhead{Ref.} \\
  \colhead{(1)} &
  \multicolumn{2}{c}{(2)} &
  \multicolumn{2}{c}{(3)} &
  \multicolumn{2}{c}{(4)} &
  \colhead{(5)}}
  
\startdata
00274197$+$0503417   &   13.24   &   1.56   &    10.3   &   0.3   &   94.8   &   1.5   &   1  \\
\nodata              &   10.4    &   0.8    &    16.7   &   1.1   &   94.8   &   3.8   &   2  \\
\\
00470038$+$6803543   &   81.49   &   1.91   &   435.6   &   1.0   &  116.8   &   0.2   &   1  \\
\nodata              &   82.3   &   1.8   &   433.7   &   1.2   &  117.5   &   0.2   &   2  \\
\\
01144035$+$4428465   &   35.02   &   0.62   &   913.9   &   0.4   &   94.7   &   0.2   &   1  \\
\nodata              &   34.10   &   3.20   &   887.6   &   1.0   &   94.9   &   0.6   &   3  \\
\\
01365662$+$0933473   &  162.13   &   0.57   &  1237.9   &   0.2   &   91.1   &   0.1   &   1  \\
\nodata              &  162.88   &   0.98   &  1222.7   &   0.8   &   90.0   &   0.1   &   4  \\
\\
01410321$+$1804502   &   41.81   &   0.73   &   404.4   &   0.3   &   97.7   &   0.1   &   1  \\
\nodata              &   44.06   &   2.05   &   408.1   &   0.1   &   96.9   &   0.1   &   5  \\
\nodata              &   41.0    &   2.8    &   412.7   &  10.0   &   96.5   &   1.4   &   6  \\
\enddata
\tablecomments{Table~\ref{tab:astcomp} is published in its entirety in machine-readable format.  A portion is shown here for guidance regarding its form and content.}
\tablerefs{(1) This paper, Table~1,
           (2) \citet{2016ApJ...833...96L},
           (3) \citet{2014ApJ...784..156D},
           (4) \citet{2016AJ....152...24W},
           (5) \citet{2014PASP..126...15W},
           (6) \citet{2016ApJS..225...10F},
           (7) \citet{2016AJ....151..160F},
           (8) \citet{1995AJ....110.3014T},
           (9) \citet{2006AJ....132.2360H},
           (10) \citet{2009AJ....137..402G},
           (11) \citet{2005AJ....130..337C},
           (12) \citet{2011AJ....141..117J},
           (13) \citet{2014A+A...568A...6Z},
           (14) \citet{1995gcts.book.....V}, Vol. II,
           (15) \citet{2014AJ....147...94D},
           (16) \citet{2004AJ....127.2948V},
           (17) \citet{2012ApJS..201...19D},
           (18) \citet{2010A+A...514A..84S},
           (19) \citet{2013MNRAS.435.1083K},
           (20) \citet{2011AJ....141...54A},
           (21) \citet{2007A+A...464..787S},
           (22) \citet{2009A+A...493L..27S},
           (23) \citet{2005AJ....129.1954J},
           (24) \citet{1996MNRAS.281..644T},
           (25) \citet{2006AJ....132.1234C},
           (26) \citet{1997AJ....114.1268M},
           (27) \citet{2012ApJ...752...56F},
           (28) \citet{2010AJ....140..897R},
           (29) \citet{2017MNRAS.466.4211G},
           (30) \citet{2009ApJ...700..623P}
}
\end{deluxetable}

%

The agreement between USNO CCD determinations and the MEarth results
for the 33 stars in common is much less satisfactory.  Figure~\ref{fig:compare}, Panel~a shows
evidence for systematic differences in the sense that the MEarth
parallaxes for this subsample are larger than our CCD determinations.
\citet{2014ApJ...784..156D} estimate that for their full 1057 star sample
they are realizing an average precision per star of 5 mas.  The
cataloged results for this 33 star subset give an average \replaced{error}{uncertainty} for
their absolute parallaxes of 4.3 mas.  However, there are 15 stars in
Figure~\ref{fig:compare} where the MEarth parallax is larger than the USNO value by
over 8 mas, and as large as 26 mas.

Figure~\ref{fig:compare}, Panel~b shows our Table~\ref{tab:ast} absolute parallax determinations compared
with the published results from the RECONS/CTIOPI Programs \citep{2005AJ....130..337C,2006AJ....132.1234C,2005AJ....129.1954J,2011AJ....141..117J,2006AJ....132.2360H,2010AJ....140..897R,2014AJ....147...94D} and the Torino
Observatory Parallax Program \citep[TOPP;][]{2007A+A...464..787S,2010A+A...514A..84S}.  We have
23 and 15 stars, respectively, in common with these two programs.
Both programs employ direct CCD imaging, RECONS/CTIO with 0.9-m and 1.5-m
reflectors and TOPP with their 1.05-m scaled-down version of NOFS's
Strand Astrometric Reflector.  No systematic trends or significant
outliers are seen in this comparison.

Parallax determinations carried out at near-infrared wavelengths
include a number of stars in common with our Table~\ref{tab:ast} sample.  Five
ultracool subdwarfs in Table~\ref{tab:ast} of this paper were also observed by
\citet[][hereafter SRS09]{2009A+A...493L..27S} at $H$-band using the 3.5-m telescope
located in Calar Alto, Spain.  \citet{2012ApJS..201...19D} and \citet{2016ApJ...833...96L} employed the
Canada-France-Hawaii Telescope (CFHT) to observe a total of 20 of our stars (10 and 10, respectively)
either at $J$-band or with a narrow $K_s$-band filter.  Finally, seven of our stars have preliminary parallaxes from the USNO ASTROCAM program being carried out at
NOFS on the Strand Astrometric Reflector \citep{2004AJ....127.2948V}.  These observations were made at either $J$-band or
$H$-band.  Figure~\ref{fig:compare}, Panel~c shows the comparison with our CCD determinations.

Two collaborations, both operating in the southern hemisphere, have
commissioned specially designed cameras for astrometrically detecting
planets around nearby late-type M, L, and T dwarfs.  A by-product of
such observations is the parallax determinations for the target stars.
One group, employing a camera with red optimized CCDs on the ESO Very
Large Telescope, has presented absolute parallaxes for 20 M8.0V to
L2.5V stars \citep{2014A+A...565A..20S}.  The precisions of these
determinations are unusually good by ground-based standards, with an
average \replaced{formal mean error}{uncertainty} of $\pm$0.10 mas.  Unfortunately, none of
these stars are in common with ours.

The second collaboration is the Carnegie Astrometric Planet Search
(CAPS) program, which employs the 2.5-m du Pont Telescope at the Las
Campanas Observatory \citep{2009PASP..121.1218B}.  Their camera employs a
HAWAII-2RG infrared array with a filter defined bandpass of $\sim100$ nm
FWHM centered at about 865 nm -- optimum for observations of M dwarfs.
\citet{2016AJ....152...24W} have presented absolute parallaxes for 134 of
their targets, of which 19 are in common with our Table~\ref{tab:ast} sample.
Except for 2MASS J14162408+1348263, which has the largest parallax error ($\pm$4.99
mas) among their entire 134 targets, their average mean error for the
other 18 stars in common with ours is $\pm$1.01 mas.  Figure~\ref{fig:compare}, Panel~d shows the
CAPS versus USNO CCD comparison where there is some indication of a small
($\sim$ 1.0 -- 1.5 mas), systematic offset between the two, in the sense that
the CAPS values are slightly smaller.  Such a small difference could
possibly be due to differences in arriving at the correction from
$\pirel$ to $\piabs$ for the two programs.  That is, the global
solutions for CAPS's so-called ``zero point offsets'' versus the mean
photometric parallaxes of the individual reference stars employed by
USNO.

Finally, Figure~\ref{fig:compare}, Panel~e shows the comparison with 11 additional parallax
determinations from 7 different sources: \citet{1995AJ....110.3014T,1996MNRAS.281..644T};
\citet{1997AJ....114.1268M}; \citet{2009AJ....137..402G}; \citet{2009ApJ...700..623P}; \citet{2012ApJ...752...56F}; \citet{2014A+A...568A...6Z}; and \citet{2016ApJS..225...10F}.  We note that for the three determinations
where the combined errors for the differences with our results are
smallest -- the {\it HST}-FG determination for the distant subdwarf G166-37
(2M1434+2510) by \citet{1997AJ....114.1268M}; the determination for
LHS 269 by \citet{2009AJ....137..402G}; and the determination for LHS 474
(vB10) by \citet{2009ApJ...700..623P} -- there is a small offset of $\sim$~1 mas
in the sense that our Table~\ref{tab:ast} parallaxes are larger.  This is similar
to what we noted for the CAPS parallax determinations in Figure~\ref{fig:compare}, Panel~d.
However, the difficulties with observing LHS 474 were noted above, and
the small systematics for the other two stars are still less than the
combined errors of the comparisons with our results.

\section{Photometric Results}
\label{section:photometry}

\begin{figure}
\plotone{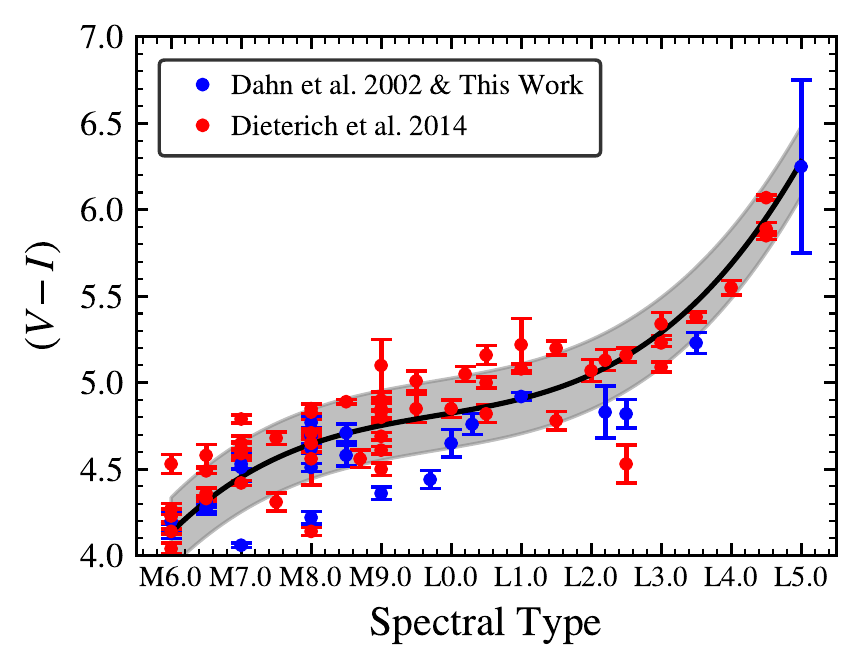}
\caption{Spectral type versus $V-I$.  The fitted curve displays the calibration used to convert spectral type to $V-I$ for stars lacking $V$-band photometry.  \added{The gray band represents the standard deviation of the residuals to the fit.}}
\label{fig:spectral-type-vi}
\end{figure}

Photometry in the $V$ and $I$ bands of the Johnson-Kron-Cousins system \citep{1976MmRAS..81...25C,1980SAAOC...1..234C} was measured for both the astrometric reference stars and
and parallax targets in most of the fields using the NOFS 1.0m
reflector.  The secondary photometric standards from \citet{1983AJ.....88..439L,1992AJ....104..340L,2007AJ....133.2502L,2013AJ....146..131L} were employed.  Further details regarding our photometric
measures can be found in \citet[][Section 3.1]{2002AJ....124.1170D}.  A few fields
were observed in the SDSS $gri$ bands \citep{1996AJ....111.1748F} using the
NOFS 1.3m telescope and transformed to $V$ and $I$ using the relations
presented by \citet{2007AJ....134..973I}.  Where the parallax targets were
too faint at $V$-band to be measured at NOFS -- and these include many
of the earlier L dwarfs, all of the L dwarfs later than L5, and all four
of the T dwarfs -- $V-I$ was estimated for stars having spectral types
between L0 and L5 from the calibration curve shown in Figure~\ref{fig:spectral-type-vi}.  This
curve was constructed using stars with measured $V$-band measures from
\citet{2014AJ....147...94D} and this study.  The third-degree polynomial fit is given by
\replaced{\begin{eqnarray}
V - I & = & 4.824 + 0.06942 I_{\rm SpT} + \\
\nonumber
        &   & 0.005344 I_{\rm SpT}^2 + 0.007779 I_{\rm SpT}^3,
\end{eqnarray}
}{\begin{eqnarray}
V - I & = & 4.824 (0.040) + \\
\nonumber
       &  & 0.06942 (0.024) I_{\rm SpT} + \\
\nonumber
        &   & 0.005344 (0.0033) I_{\rm SpT}^2 + \\
\nonumber
        &  & 0.007779 (0.0015) I_{\rm SpT}^3,
\end{eqnarray}}
where $I_{\rm SpT}$ is an index with zero at spectral type L0, and incremented by numerical spectral subtype.  \added{The fit has a standard deviation of 0.20 mag}.  Thus, an L1.0 star has $I_{\rm SpT} = 1.0$, while an M9.0 star has $I_{\rm SpT} = -1.0$.
This fit should not be extrapolated beyond the boundaries indicated in Figure~\ref{fig:spectral-type-vi}.  The $V-I$ values so derived
were only employed for the color term in the reduction of actual
$I$-band observations.

The $VI$ photometry so obtained was employed for three purposes.
First, the $V-I$ color was used to derive DCR corrections for
observations taken off of the meridian.  However, ever since the
TEK2K and EEV24 cameras were commissioned, our policy has been to
center each exposure within $\pm$15 minutes of meridian crossing.  A few
frames with exposures centered as much as $\pm$30 minutes off the
meridian have been retained and used.  Hence, DCR is very much a
second-order correction for our parallax determinations from these two
cameras.

\begin{figure}
\plotone{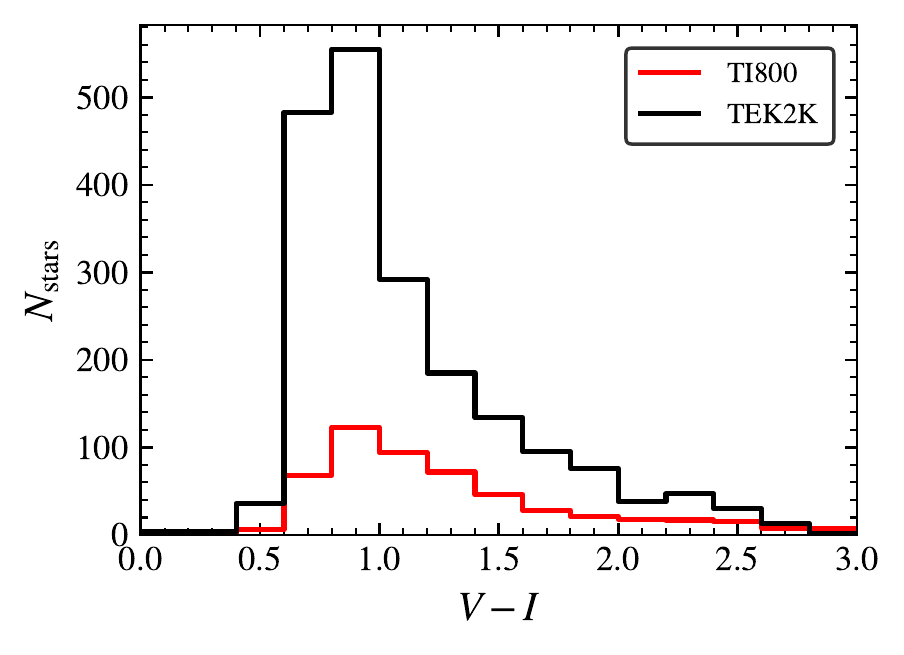}
\caption{Distribution of reference stars in $V-I$ color, shown separately for the TI800 and TEK2K programs.}
\label{fig:rsVmI}
\end{figure}

Second, our $VI$ photometry for the reference stars in each of the
parallax target fields is used in the selection of which stars
actually get employed in the astrometric reductions.  The goal here is
to avoid using stars that might be nearby, and to that end, whenever
possible, red stars are (1) rejected as potentially being nearby
dwarfs or (2) only employed if they appear to be distant giants.
Figure~\ref{fig:rsVmI} shows the frequency distribution for the 1992
individual reference stars in all of the TEK2K solutions and for the
517 individual stars employed in all of the TI800 solutions.  For
the TEK2K fields, the $V-I$ colors peak strongly in the $0.6 < V-I <
1.0$ range.  Since the reference frame stars that we are employing are
faint -- mostly in the $16<V<18$ range -- our calibration for photometric
parallax \citep[][Figure 1]{2016AJ....152..118H} places them at distances of 
generally 1.0--2.0 kpc.  Reddening/extinction can be important for fields
near the galactic plane, and can be handled very satisfactorily as explained
in \citet[][their Section 3.2]{2016AJ....152..118H}.

Third, our photometry for the parallax stars themselves is
employed for the color versus absolute magnitude and color versus color
diagrams presented in Section 5 below for astrophysical interpretation.
The parallax stars are mostly at distances less than 100 pc, and hence,
within the so-called ``Local Bubble'' \citep{1968AJ.....73..983F,2011ARA&A..49..237F}.  This a volume around the sun wherein, even in the
galactic plane, extinction by interstellar dust is minimal,
and which extends hundreds of parsec in the directions toward the
galactic poles.  However, in a few directions -- such as for our three
fields toward the Hyades cluster -- it can still be an issue.

Table~\ref{tab:pht} presents our NOFS photometric results for each of the parallax
stars in Table~\ref{tab:ast}.  Column (1) is the same as in Table~\ref{tab:astcomp}.
Columns (2) and (3) are the NOFS $V$ and $I$ measures along with
their formal uncertainties, all rounded to the nearest 0.01 mag.  The
results quoted for $V$ and $I$ include measures previously reported in
\citet{1992AJ....103..638M} and in \citet{2002AJ....124.1170D}.  Hence, the Table~\ref{tab:pht}
values supersede the earlier ones.  Where the star was too faint to be
measured at $V$-band, and the spectrally determined $V-I$ was employed for
the color term in the $I$-band reductions (see above), the quoted \replaced{error}{uncertainty}
for $I$ has been adjusted upward appropriately.  And since Figure~\ref{fig:spectral-type-vi}
could neither be calibrated for L dwarfs/subdwarfs later than L5 nor for T
dwarfs, further uncertainties in the reported $I$ magnitudes are flagged
with a colon.

\begin{deluxetable*}{cr@{$\pm$}rr@{$\pm$}rr@{$\pm$}rr@{$\pm$}rr@{$\pm$}rcc}
\tabletypesize{\small}
\tablewidth{0pt}
\tablenum{3}
\tablecaption{Photometric Results
\label{tab:pht}}
\tablehead{
	\colhead{Name\tablenotemark{a}} & 
	\multicolumn{2}{c}{$V$} &
	\multicolumn{2}{c}{$I$} &
	\multicolumn{2}{c}{$J$} &
	\multicolumn{2}{c}{$H$} &
	\multicolumn{2}{c}{$K_S$} &
	\colhead{SpT} & 
	\colhead{Type\tablenotemark{b}}   \\*
	\colhead{(1)}  &
	\multicolumn{2}{c}{(2)}  &
	\multicolumn{2}{c}{(3)}  &
	\multicolumn{2}{c}{(4)}  &
	\multicolumn{2}{c}{(5)}  &
	\multicolumn{2}{c}{(6)}  &
	\colhead{(7)}  &
	\colhead{(8)}}
\startdata
 \object[2MASS J00110078+0420245]{00110078$+$0420245}  &   17.89   &   0.02   &   15.57   &   0.02   &   14.34   &   0.03   &   13.81   &   0.05   &   13.76   &   0.04    &       usdM4.5    &         5  \\
 \object[2MASS J00192745+0450297]{00192745$+$0450297}  &   17.25   &   0.02   &   13.75   &   0.02   &   11.98   &   0.03   &   11.40   &   0.03   &   11.09   &   0.02    &         M5.5V    &         1  \\
 \object[2MASS J00274197+0503417]{00274197$+$0503417}  &   \multicolumn{2}{c}{\nodata}   &   19.79   &   0.10   &   16.19   &   0.09   &   15.29   &   0.10   &   14.96   &   0.12    &      M9.5pecV    &         1  \\
 \object[2MASS J00312326+0936169]{00312326$+$0936169}  &   16.56   &   0.02   &   14.72   &   0.02   &   13.64   &   0.02   &   13.10   &   0.03   &   12.91   &   0.03    &       \nodata    &       3:v  \\
 \object[2MASS J00350768+7627544]{00350768$+$7627544}  &   17.78   &   0.02   &   13.88   &   0.02   &   11.65   &   0.02   &   11.06   &   0.03   &   10.70   &   0.03    &       \nodata    &         1  \\
 \object[2MASS J00361617+1821104]{00361617$+$1821104}  &   21.34   &   0.06   &   16.11   &   0.01   &   12.47   &   0.03   &   11.59   &   0.03   &   11.06   &   0.02    &          L3.5    &         6  \\
 \object[2MASS J00470038+6803543]{00470038$+$6803543}  &  \multicolumn{2}{c}{\nodata}  &   19.77   &   0.12:   &   15.60   &   0.07   &   13.97   &   0.04   &   13.05   &   0.03    &         L7pec    &         6  \\
 \object[2MASS J00475502+4744342]{00475502$+$4744342}  &   16.00   &   0.05   &   13.20   &   0.05   &   11.68   &   0.02   &   11.17   &   0.02   &   10.91   &   0.02    &       \nodata    &       2:l  \\
 \object[2MASS J00510351-1411047]{00510351$-$1411047}  &   19.82   &   0.04   &   16.12   &   0.03   &   14.09   &   0.03   &   13.53   &   0.02   &   13.32   &   0.03    &       \nodata    &       3:v  \\
 \object[2MASS J00554418+2506235]{00554418$+$2506235}  &   18.94   &   0.03   &   15.76   &   0.02   &   14.26   &   0.03   &   13.78   &   0.03   &   13.57   &   0.04    &        sdM6.5    &      3:vs  \\
\enddata
\tablecomments{Table~\ref{tab:pht} is published in its entirety in machine-readable format.  A portion is shown here for guidance regarding its form and content.}
\tablenotetext{a}{Name corresponds to the ``2MASS J'' identifier in all but five cases where the LHS name is used.}
\tablenotetext{b}{Our adopted types are as follows: (1) K or M Dwarf;
  (2) an uncertain subdwarf with 150 $\le$ V$_{\rm tan} <$ 180 km~s$^{-1}$
  and/or showing marginal subluminosity in one or more of the diagrams
  presented in Section~\ref{sec-diagrams}; (3) K or M subdwarf (but not extreme nor
  ultra); (4) exteme M subdwarf (esdM); (5) ultra M subdwarf (usdM); (6)
  L or T dwarf (d); (7) L subdwarf (sdL); (8) extreme L subdwarf (esdL);
  (9) ultra L subdwarf (usdL).  Type assignments for 3 -- 9 are based
  on observed spectroscopic classifications found in the literature.
  For types 2 and 3, a following letter indicates the supporting
  evidence for our typing decision where: v = V$_{\rm tan} \ge$ 180 km
  s$^{\rm -1}$; s = spectral type; and l = significant subluminosity
  in one or more of color-absolute magnitude diagrams presented in
  Section~\ref{sec-diagrams}.}

\end{deluxetable*}


Columns (4) -- (6) in Table~\ref{tab:pht} give the 2MASS $JHK_s$
magnitudes extracted from the Point Source Catalog \citep{2006AJ....131.1163S} and again rounded to the nearest 0.01 mag.

Column (7) of Table~\ref{tab:pht} gives, where available, representable
spectral types extracted from the literature.  Of the total 309 stars,
spectral typings were located for 245 ($\sim$~80\%).  (We do not consider the
Luyten color classes -- such as k, k-m, m and m+ -- as legitimate
spectral types since they merely represent broadband photographically
determined colors with significant measurement uncertainties.)  The
spectral types that we adopt are not on any homogeneous system, and we
offer no attempt to do so.  Of the 245 stars for which we quote types,
the majority ($\sim$~75\%) are M stars -- either dwarfs or subdwarfs.  The
remaining include (disregarding several faint, unresolved secondary
companions in binary systems) 6 late K dwarfs or subdwarfs, 43 L
dwarfs, 6 L subdwarfs, and 4 T dwarfs.

For the M dwarfs, we give priority to typings on the system set up by \citet{1991ApJS...77..417K}.  As for M subdwarfs, \citet{2008AJ....136..840J} have
presented an excellent summary of the evolution of K-M subdwarf spectral
typings.  Using multichannel spectrophotometric observations of a
large sample of cool (late-type) proper motion stars, \citet{1980ApJ...240..859A} identified four stars -- LHS 453, LHS 3382, G 7-17 and LHS 498 --
for which the metallicity indicators TiO, CaH, MgH, and CaI
$\lambda$4227\AA\ indicated extreme metal deficiency.  \citet{1984ApJ...286..269H} used photometric and spectroscopic observations to
establish narrowband indices measuring the relative strengths of the
TiO and CaH features in the $\lambda\lambda$6800--7600\AA\
wavelength range.  \citet{1997AJ....113..806G} and \citet{1997PASP..109.1233G,1997PASP..109..849G}
followed with a similar scheme that employs three indices for
measuring the strength of CaH along with a single index measuring the
strength of TiO in the $\lambda\lambda$6200--7400\AA\ window.  This
enabled M dwarfs to be identified as either dwarfs, subdwarfs (sdM) or
extreme subdwarfs (esdM).  As originally defined, this system only
covered earlier M subdwarfs, down to $\sim$~M5.  Among the very coolest
subdwarfs, LHS 377 was set to {\it define} the type sdM7.0 \citep{2006AJ....132.2372G} since it was the only subdwarf later than $\sim$~sdM5 known at
that time.  Extension to cooler subdwarf M stars was carried out by
\citet{1999A+A...350L..62S}, \citet{2003AJ....125.1598L}, \citet{2004ApJ...602L.125L},
and \citet{2004A+A...425..519S}.  \citet{2007ApJ...669.1235L} then proposed a
revision to this classification scheme by introducing an additional
parameter that measures the strength of the TiO compared with a
similar star possessing solar metallicity.  This resulted in
establishing an additional subclass of ``ultrasubdwarf'' corresponding
to subdwarfs with the lowest metal content.  \citet[][Table 2]{2007ApJ...669.1235L} presented a listing of classification standard stars for all
subclasses sd, esd, and usd.  In our Table~\ref{tab:ast} we present parallax
determinations for 11 of these stars -- LHS 228 (sdM2.0); LSR 1425+7102
(sdM8.0); LHS 360 (esdM0.0); LHS 2045 (esdM5.0); LEHPM 2-59 (esdM8.5);
LHS 2843 (usdM0.0); LHS 325 (usdM3.5); LHS 1032 (usdM4.5); LHS 2500
(usdM5.0); LSPM J0822+1700 (usdM7.5); and 2MASS J12270506-0447207 (usdM8.5).

\citet{2008AJ....136..840J} pointed out that this current system is somewhat
deficient in that it utilizes a rather limited region of the spectrum
at low resolution ($\lambda/\Delta\lambda \sim$ 3,000) and does not
adequately link the low-metallicity subdwarfs to their main-sequence
counterparts with respect to temperature and surface gravity.  \citet{2009PASP..121..117W} argue that the gravity will only be an issue for young
stars that have not yet fully settled onto the main sequence.  Using
high-dispersion spectra covering $\lambda\lambda$6200--9800\AA\
($\lambda/\Delta\lambda \sim$ 33,000, resolution $\sim3.1$~\AA) to
determine Fe and Ti abundances, they demonstrate that the
low-resolution CaH and TiO indices do measure valid metallicity
discrimination among the sd, esd, and usd subclasses for stars with
$3500\ {\rm K} < T_{\rm eff} < 4000\ {\rm K}$ and with
$-1.5 < {\rm [Fe/H]} < +0.05$.  Cooler
subdwarfs (e.g., the ultracool ones later than M7) are too faint and are
not currently observable with similar high-resolution spectroscopy.

\citet{2017MNRAS.464.3040Z} have now extended the sd, esd, and usd subclasses
to L-type stars.  Optical spectra in the
$\lambda\lambda$7300--8800\AA\ wavelength region were employed to
tie the L-subdwarfs to the L-dwarf spectral standard stars \citep{2007ApJ...657..494B, 1999ApJ...519..802K,2010ApJS..190..100K}.  Although molecular bands at
optical wavelengths (e.g., CaH, TiO, VO, and FeH) remain important in
establishing the L-subdwarf subclasses, near-infrared spectral
behavior out to $\sim2.5$ microns become very important.  The defining
characteristics for the L-subdwarf subclasses are outlined in Table~3
of \citet{2017MNRAS.464.3040Z}.

Three of the stars in our sample have now been reclassified into this new
system.  SSSPM J1013-1356 was previously classified as sdM9.5 but now is
classified as usdL0,\\
2MASS J1626034+3925190 was formerly classified as sdL4
but now becomes usdL4, and LSPM J1826+3014 is now typed sdL0 rather than
d/sdM8.5.  Further modifications to this system can be expected as the
sample of L-subdwarfs grow.

The final column in Table~\ref{tab:pht} indicates the star ``type'' that we adopt
for each target based on the aggregate astrometric, photometric,
spectroscopic, and kinematic information available to us.  This ``type''
is meant to indicate whether each star is most likely to be a dwarf,
subdwarf, extreme subdwarf, or ultrasubdwarf and will be employed for
to present the diagrams in Section 5.

\section{Sample Color-Absolute Magnitude and Color-Color Diagrams}
\label{sec-diagrams}

\begin{figure*}
\plotone{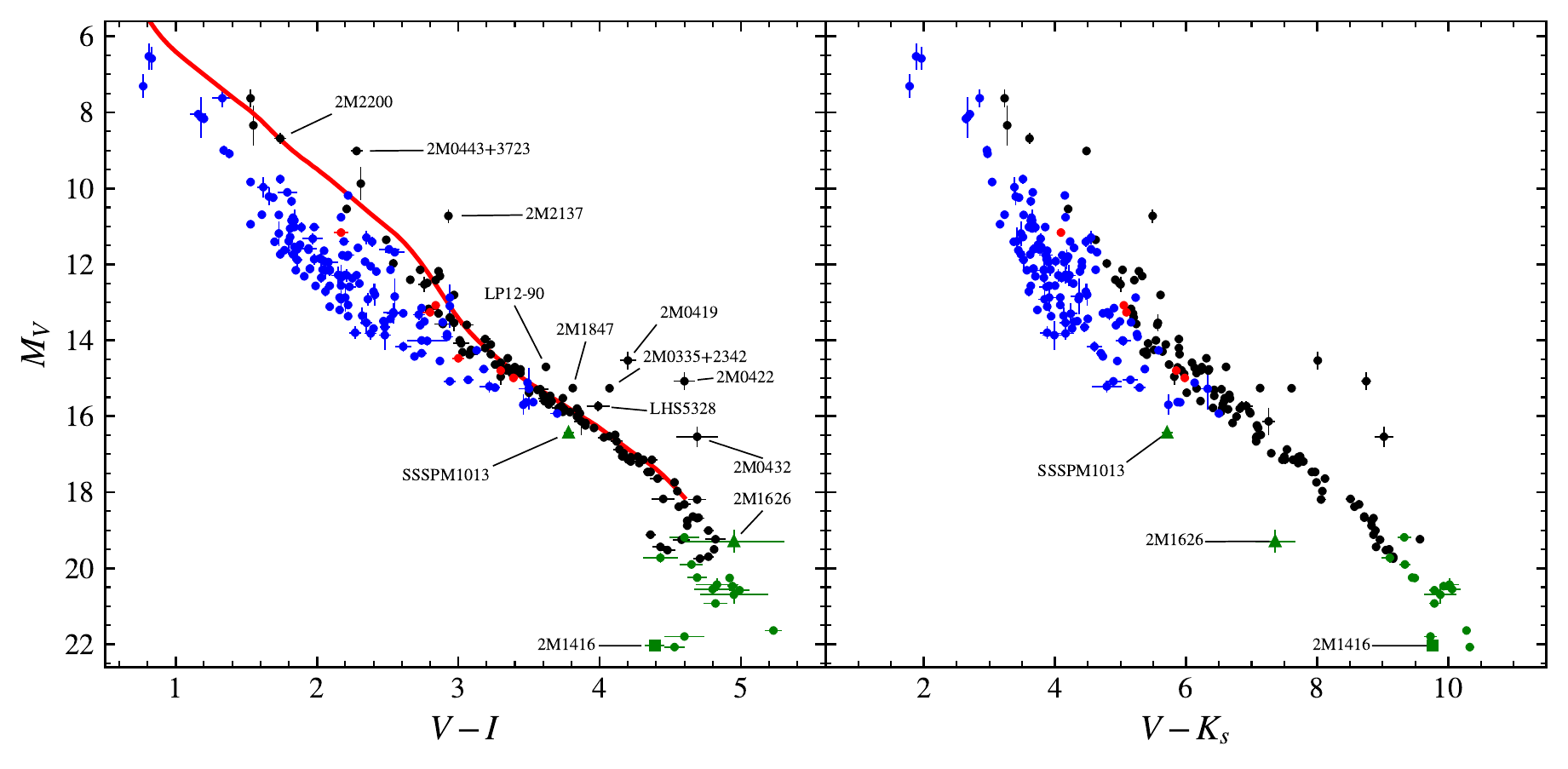}
\caption{$M_V$ versus $V-I$ (left panel) and $V-K_s$ (right panel) for all stars with $VI$ photometry.  Symbols are: K and M dwarfs (black circles), M subdwarfs (blue circles), dL (green circles), \replaced{esdL}{sdL} (green squares), usdL (green triangles), unknown dwarf/subdwarf status (red circles).  The red line is the nominal dwarf main sequence from \citet{2016AJ....152..118H}.}
\label{fig:mv}
\end{figure*}

Tables 1 and 3 above provide the basis for a large number of empirical color-absolute magnitude and color-color diagrams.  Only a sample will be presented here.
Figure~\ref{fig:mv} shows $M_V$ versus $V-I$ (left panel) and $M_V$ versus $V-K_s$ (right panel)
for the entire sample with observed $VI$ photometry in Table 3.
The filled black circles are K and M dwarfs.  The filled blue circles
are M subdwarfs (sdM, esdM, and usdM types).  The green symbols are all L dwarfs,
where the solid circles are dL types, the solid square is the \replaced{only esdL}{sdL} star
2MASS J14162408+1348263\deleted{ in our sample}, and the filled triangles are the two usdL stars \added{in our sample},
SSSPM J1013-1356 and 2MASS J1626034+3925190.  The red line is the nominal dwarf main sequence as
presented in \citet{2016AJ....152..118H}.  Here (and in all subsequent figures), formal
$\pm1\sigma$ error bars are included for all data points plotted.

One general feature of note seen in $M_V$ versus $V-I$ is a hook back to bluer colors for
the latest M-dwarfs.  This is recognized to be an effect of the change in spectral
energy distribution (primarily changes in the strength of various prominent TiO
absorption bands) across the $V$ and $I$ bandpasses as the sequence progresses to cooler
temperatures.  No such deviation is
seen in the $M_V$ vs $V-K_s$ diagram.

A few individual stars warrant special attention as obvious outliers.
SSSPM J1013-1356 has recently been
reclassified to spectral type usdL0 \citep{2017MNRAS.464.3040Z} versus the earlier
classification of sdM9.5 originally announced by \citet{2004A+A...425..519S}.  This
star's location in Figure~\ref{fig:mv} is somewhat difficult to understand
as an L star since it is roughly an order of magnitude more luminous than any
of the other L dwarfs and subdwarfs in the plot.  Clearly, the problem is not
with the astrometry.  Our Table~1 parallax is in very good agreement with the
independent determination by \citet{2009A+A...493L..27S}.  Likewise, the photometry
is believed to be reliable to the precision reported in Table 3.  A spectral
classification of very late esdM or usdM would clearly be more compitable with
our results.

2MASS J16262034+3925190 was first announced as sdL4 by \citet{2004ApJ...614L..73B}.  \citet{2015A+A...579A..58L} observed an upper limit of 90~m\AA\ for LiI (6707\AA) absorption, placing
it in the Li-burning regime.  They also measured an improved $V_{\rm rad}=239\pm12~\kmsec$
which, along with our $V_{\rm tan}=212~\kmsec$, strongly supports halo
candidacy.  This star stands out in the $M_V$ vs $V-K_s$ diagram but does not stand out in
$M_V$ vs $V-I$.  As we noted above in Section~\ref{section:unresolved-binaries}, the residuals from our parallax
astrometry suggest a low-amplitude perturbation.

The three dwarfs located in the direction of the Hyades cluster, but apparently
well beyond it, -- 2MASS J04193967+1433329, 2MASS J04223075+1526310, and\\ 2MASS J04325119+1730092 -- fall significantly
above the dwarf sequence, as do two of the three moving group candidates\\
(2MASS J04435686+3723033 and \\2MASS J21374019+0137137).  As noted in Sec. 3 above, images of 2MASS J21374019+0137137
taken in very good seeing suggest a partially resolved binary system.  Also falling
above the dwarf sequence are (1) 2MASS J03350208+2342356, an M8.5V star with detected LiI
absorption, consistent with pre-main-sequence status, and (2) stars LP12-90,
2MASS J18470342+5522433, and LHS 5328 which appear to be unresolved binaries with nearly equal-luminosity components.

\begin{figure}
\plotone{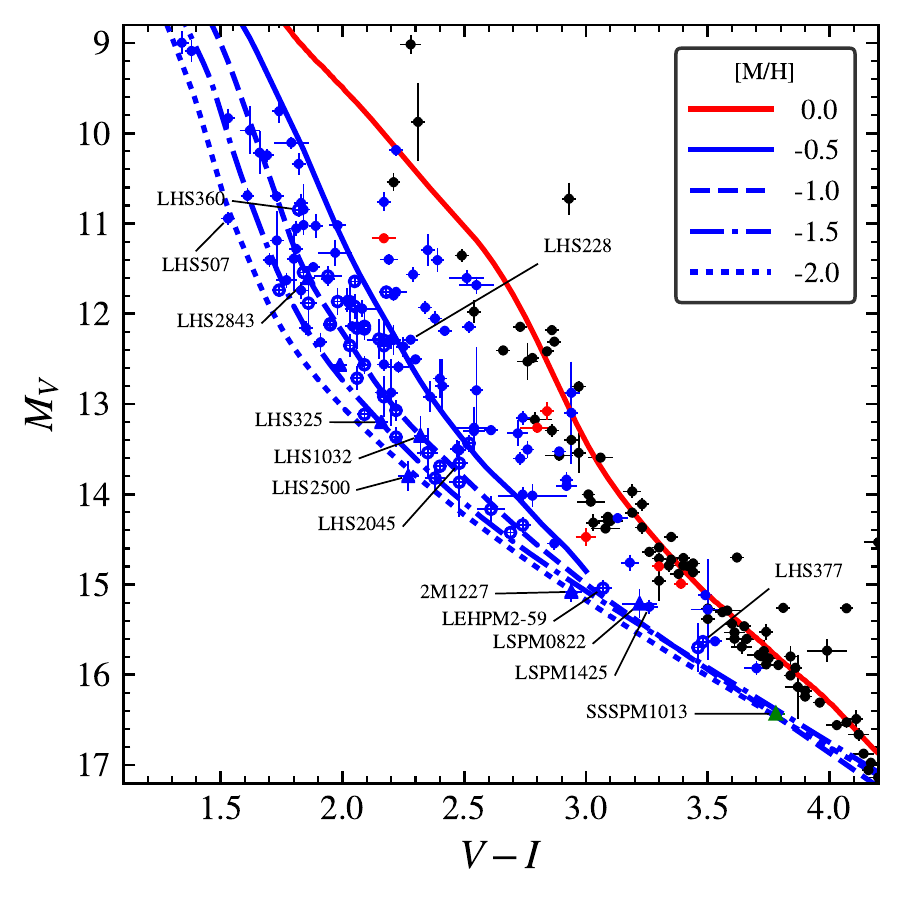}
\caption{$M_V$ versus $V-I$ for all stars with $VI$ photometry, zoomed in on the K and M subdwarf region.  Symbols are: K and M dwarfs (filled black circles), sdK and sdM (filled blue circles), dwarf/subdwarf status unknown (filled red circles), esdM (open blue circles), usdM (filled blue triangles), SSSPM J1013-1356 (green triangle).  The red line is the nominal dwarf main sequence from \citet{2016AJ....152..118H}.  Blue lines are model locii for metal poor stars from \citet{1997A+A...327.1054B, 1998A+A...337..403B}.}
\label{fig:mv-vi-zoom}
\end{figure}

Figure~\ref{fig:mv-vi-zoom} shows an expanded view of the K and M subdwarf region of the $M_V$ vs
$V-I$ plane.  Again, K and M dwarfs are \replaced{respresented}{represented} by solid black circles, while
the sdK and sdM stars are shown as filled blue circles.  The solid red circles
are stars for which the dwarf versus subdwarf status is uncertain.  The open
blue circles are esdM stars and the solid filled blue triangles are usdM stars.
Individual stars labeled include the 11 spectral classification standards from
\citet[][Table 2]{2007ApJ...669.1235L} and LHS 377, the sdM7.0 standard designated by
\citet{2006AJ....132.2372G} but reclassified as esdM7.0 by \citet{2017MNRAS.464.3040Z}.
The solid blue lines are model locii for metal-poor stars from \citet{1997A+A...327.1054B, 1998A+A...337..403B}.  Also labeled is LHS 507 which was independently identified to be
a late subdwarf K star by \citet{1979ApJ...227..232B} using broadband
photometry and by \citet{1979ApJ...233..226L} using low-resolution spectroscopy.
The location of LHS 507 in Figure~\ref{fig:mv-vi-zoom} indicates $[{\rm M/H}]=-2.0$ such that modern
spectroscopy should confirm it to be of usdK type.  The location of SSSPM J1013-1356 is
represented by a solid green triangle.

Comparing our Figure~\ref{fig:mv-vi-zoom} observational results with \citet{1997A+A...327.1054B, 1998A+A...337..403B}
models shows no evidence for stars more metal poor than $[{\rm M}/{\rm H}]=-2.0$.  Further
spectroscopic observations of many of the stars only identified as subdwarfs
based on location in our diagrams and/or large measured $\vtan$ values (e.g., LHS 335,
LHS 1894, LHS 1953, LHS 2929, LHS 3207) should establish additional esdM- and usdM-type stars.
      
\begin{figure*}
\plotone{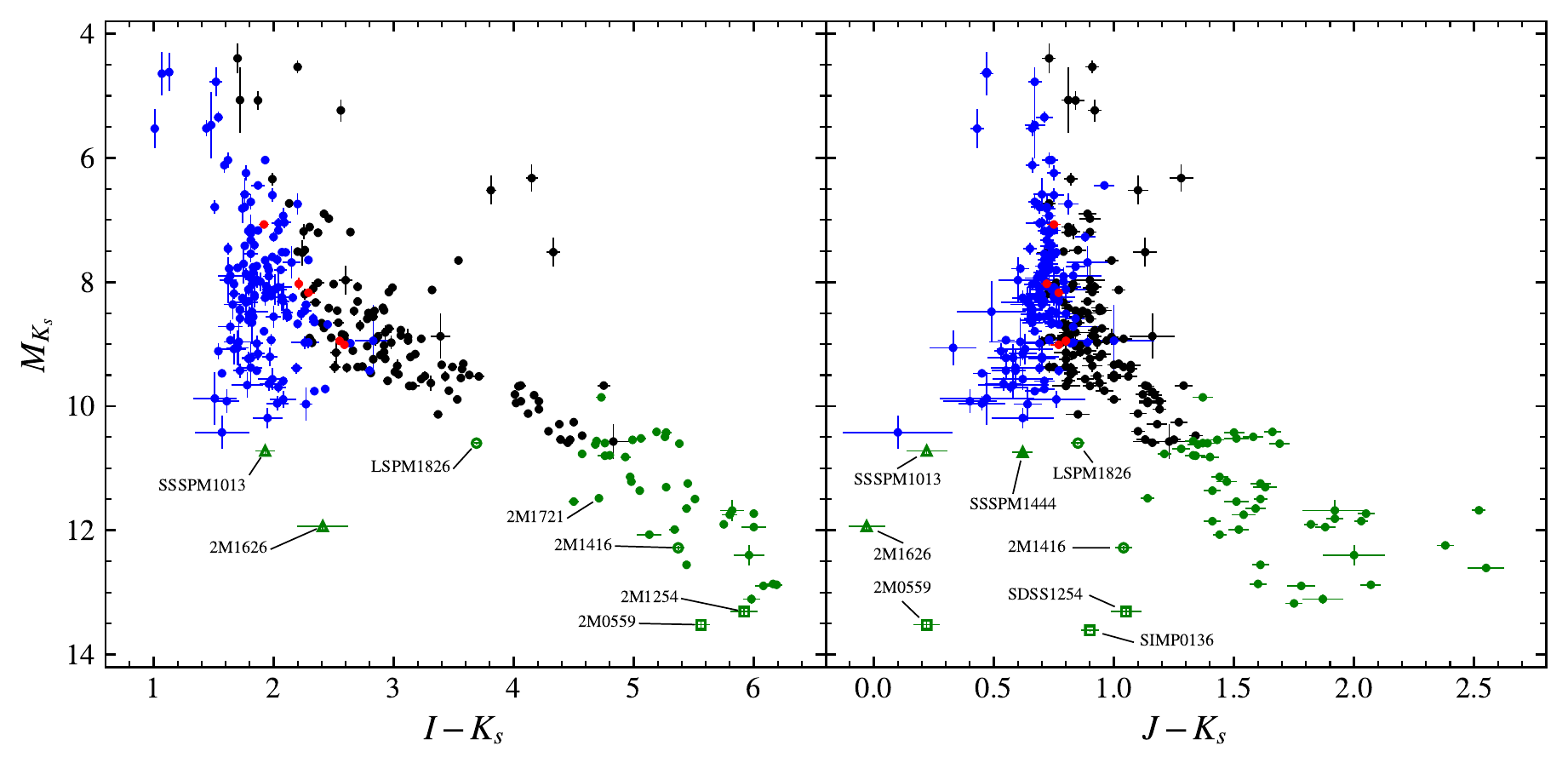}
\caption{$M_{K_s}$ versus $I-K_s$ (left panel) and $J-K_s$ (right panel) for our sample, zoomed in on the L dwarfs.  Symbols are: K and M dwarfs (filled black circles), sdK and sdM (filled blue circles), dL (filled green circles), sdL (open green circles), usdL (open green triangles), T dwarfs (open green squares).}
\label{fig:mk}
\end{figure*}

Figure~\ref{fig:mk} shows our sample in $M_{K_s}$ versus $I-K_s$ (left panel) and $M_{K_s}$ versus $J-K_s$ (right panel), focusing on the L dwarfs.
Here, the dL stars are represented by filled green circles, the sdL stars by open
green circles, the usdL stars by open green triangles, and the two T dwarfs with
$I$-band photometry (2MASS J12545393-0122474 and 2MASS J05591914-1404488) by open green squares.  The 
single esdL star in our sample (SSSPM J1444-2019) lacks $I$-band photometry.

\begin{figure}
\plotone{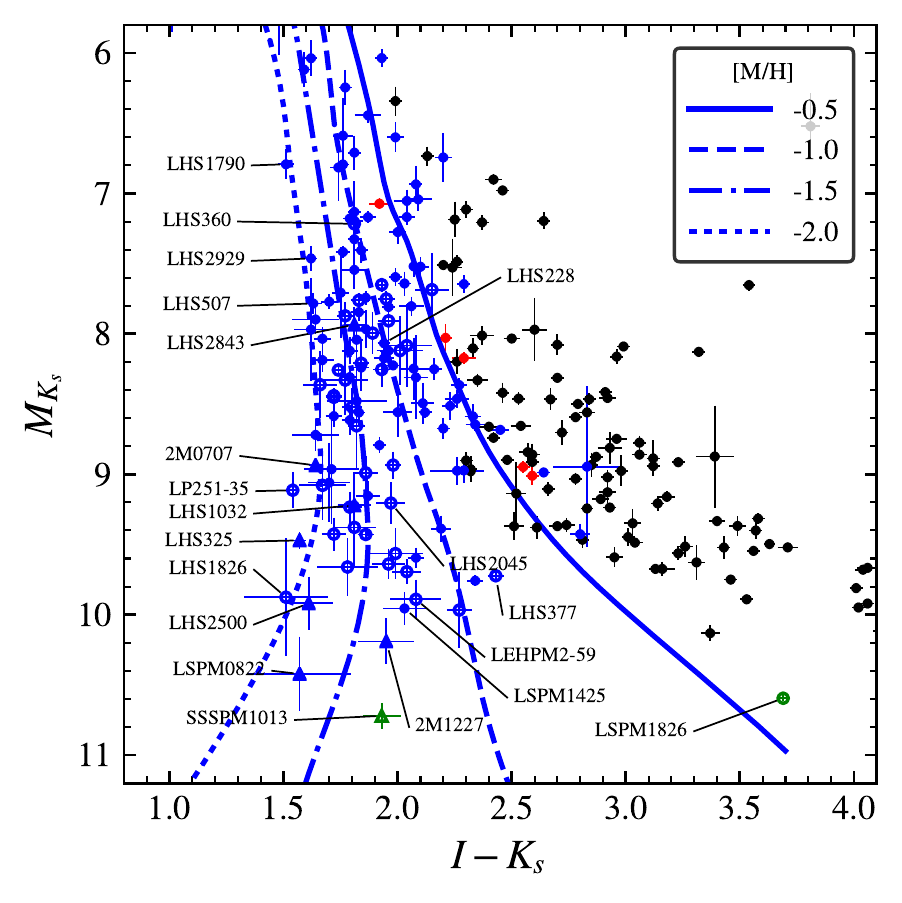}
\caption{$M_{K_s}$ versus $I-K_s$, zoomed in on the K and M dwarf region.  Symbols are the same as for Figure~\ref{fig:mv-vi-zoom}, with the following additional symbols: sdL (open green circles), usdL (open green triangles).}
\label{fig:mk-ik-zoom}
\end{figure}

Figure~\ref{fig:mk-ik-zoom} presents an expanded view of the K and M dwarf region of the $M_{K_s}$ 
versus $I-K_s$ plane.  Symbols employed are as described for Figure~\ref{fig:mv-vi-zoom}.  Additionally, sdL stars are represented by open green circles and usdL stars by open green triangles.  Again, there
seems to be many sdM stars awaiting confirmation as esdM and usdM types.  As in
Figure~\ref{fig:mv-vi-zoom}, there is scant evidence for stars with $[{\rm M}/{\rm H}]<-2.0$.

These color-magnitude diagrams using IR colors can provide an excellent
\replaced{diagostic}{diagnostic} for metallicity, and thus for subluminosity
and for stellar population membership.  They are independent of spectral
classification, and so will be useful in interpreting the
results of {\it Gaia} parallaxes for many late-type dwarfs
soon to be available (see Section~\ref{sec:discussion} below).  Early versions
of these diagrams were shown by \citet{2015csss...18..959D}.

\begin{figure}
\plotone{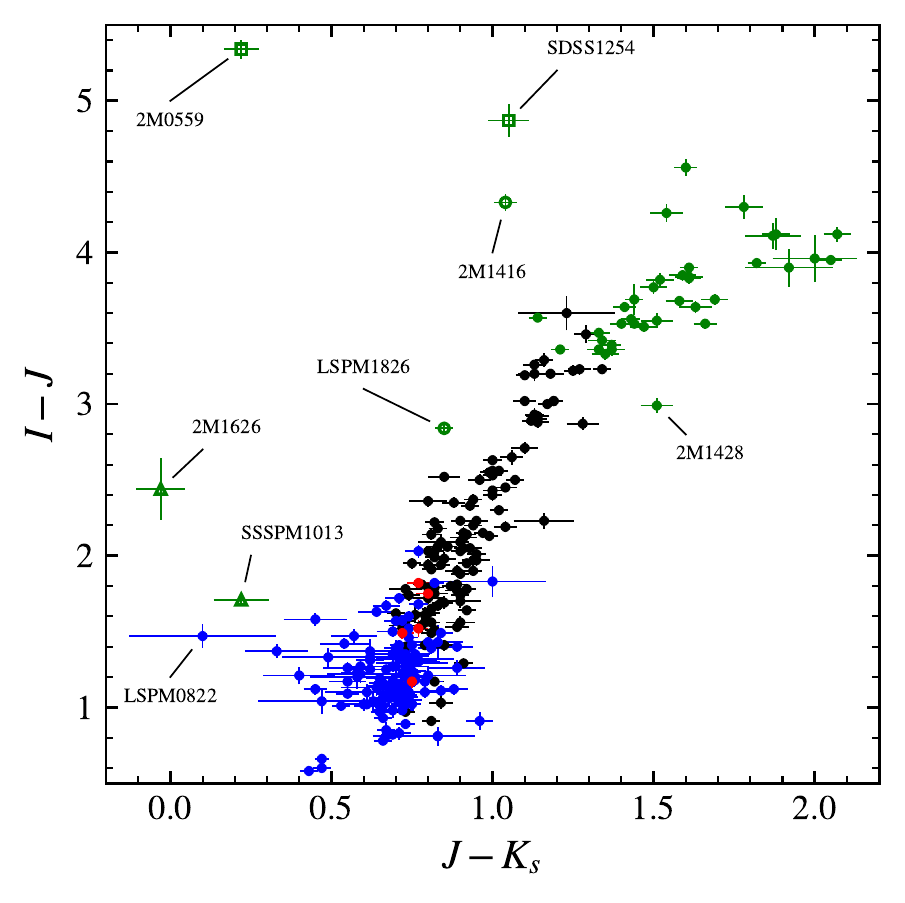}
\caption{\added{$I-J$ versus $J-K_s$ color-color diagram.  Symbols are the same as for Figure~\ref{fig:mk}}}
\label{fig:ij-jk}
\end{figure}

\added{Figure~\ref{fig:ij-jk} shows the $I-J$ versus $J-K_s$ color-color diagram.
The hook toward bluer $J-K$ colors at the top of the diagram
shows the onset of methane absorption in the $K$-band for T dwarfs
and the coolest L dwarfs.  The other outliers toward blue
$J-K$ colors are the extreme subdwarfs identified in previous
figures.  \citet{2017MNRAS.464.3040Z} recently employed an equivalent version of this diagram (using $J-K_s$ versus $i-J$, where $i$ is the SDSS magnitude) to identify six new L-type subdwarfs.}

\section{Discussion}
\label{sec:discussion}

\replaced{The parallax determinations reported in the present paper are strongly focused
on M subdwarfs and very late M dwarfs.  The intent was to provide a large enough
sample to facilitate understanding the physical causes for the spread in luminosity
amongst these stars (cf. Figure~\ref{fig:mv}).
Jao et al. (2008) employed then available model grids computed for
Gaia (Brott \& Hauschildt 2005) and compared these models to spectroscopic
data.  The authors suggest that the wavelength regions used, namely that which
includes the CaH and TiO bands, to establish the sd, esd, and usd metallicity
subclasses are also sensitive to surface gravity and do not accurately reflect
metallicity alone.  Thus, more work is needed for a comprehensive
understanding.}{
The parallax determinations reported in the present paper are intended to sample
(to the extent practical) a full range of metal-poor very-late-type stars.
The goals were to document the spread in luminosity among these stars
(cf. Figure~\ref{fig:mv}), to utilize color-magnitude diagrams
to classify a star's metallicity and population (cf. Figure~\ref{fig:mv-vi-zoom}),
and to facilitate understanding of the physical causes affecting their colors and
spectral types.  Spectroscopic classification for late-M and L subdwarfs
is still being developed.  For example, \citet{2008AJ....136..840J}
suggest that the wavelength regions using the CaH and TiO bands to establish the
sd, esd, and usd metallicity subclasses are also sensitive to surface gravity
and do not accurately reflect metallicity alone.  Thus, more work is needed
in spectrophotometric modeling for a comprehensive understanding.  Here, we
suggest an alternative path toward classification independent of spectral type;
based on color and parallax measurements, it can be applied to the large sample
of objects that will soon have these data from {\it Gaia} and LSST.}
      
\deleted{Gaia is a cornerstone mission in the science programme of the European
Space Agency  (ESA; Gaia Collaboration et al. 2016a,b) that will measure highly accurate parallaxes for large numbers of M stars --
both dwarfs and a large variety of subdwarfs.  For a limiting Gaia apparent
magnitude of $G \sim 20$, the sky-averaged, end-of-mission parallax error is projected
to be $\pm0.3$ mas (de Bruijne 2014).  Employing the $G$ versus $V$,$V-I$ cubic polynomial
relation from Jordi et al. (2010, Table~6) -- updated with the coefficients from de Bruijne (2014) -- we estimate that for a very late M-dwarf (typically with
$M_V \sim 19$ and $V-I \sim 4.7$; cf. Figure~\ref{fig:mv}) Gaia will reach to $V \sim 23.2$ corresponding to a
distance of roughly 70 pc.  Since this is within the local bubble, a volume of
$\sim1.4 \times 10^6\ {\rm pc}^3$ will be sampled and, adopting the local luminosity function from
Just et al. (2015), Gaia should measure parallaxes for $\sim$ 1700 $M_V \sim 19$ dwarfs, all
to an accuracy of 2\% or better.  Similar estimates for the earlier-type (M7 \& M8)
ultracool dwarfs indicates that Gaia will observe them, on average, to distances of 
110 pc and 90 pc, respectively.  Again employing the Just et al. (2015) results
presented in their Figure 6, we estimate that roughly 25000 of the M7 \& M8
ultracool dwarfs will be measured.
Thus, the orders-of-magnitude
greater sample size and exquisite accuracy of the trigonometric
parallaxes will greatly enhance our understanding of the physical
nature of these objects.}

A decade ago, there were only sixteen spectroscopically confirmed ultracool
subdwarfs \citep{2007ApJ...657..494B}.  Of these, all but two (2M0937+2931 and
2M0532+8246) had spectral types between M7 and L4. Shortly thereafter \citet{2008ApJ...681L..33L} announced the identification of 23 additional ultracool
M-type subdwarfs from SDSS DR6 spectra.  \deleted{These additions are mostly very
faint with estimated Gaia magnitudes in the 19.0 to 20.2 range.  Hence, while
Gaia will measure parallaxes for the majority of them with formal errors in
the $\pm 0.2$ -- 0.4 mas range, we predict that they will be so far away
that the quality of the results ($\sigma_{\pi} / \pi$) will be low.}
More recently, \citet{2017A&A...598A..92L} presented M-type subdwarf identifications
from their cross-matching of 2MASS, SDSS, and UKIDSS databases.  Included are
a total of 27 M-type ultracool subdwarfs (types 7.0 or later) plus three early-type
sdL stars. \deleted{These are again mostly faint objects, many with large uncertainties
for the SDSS $g$-band photometric measures.  Hence, the estimated Gaia magnitudes
calculated are equally uncertain, but it seems that roughly half will be brighter
than $G=21.0$.   However, the estimated spectroscopic distances are large (and
uncertain!) with only three (SDSS J0830+3612, SDSS J0909+1941, and SDSS J1512-0112)
estimated to lie within 100 pc of the sun.}
Since then a considerable number of additions have been identified, including
SDSS J010448.46+153501.8 which \citet{2017MNRAS.468..261Z} claim as the ``the most
metal-poor substellar object'' currently known with $[{\rm Fe/H}]=-2.4\pm0.2$ and a spectral
type usdL1.5.  \replaced{Although the additions bring the total number of ultracool subdwarfs
up to around 60, no additional M subdwarfs have been identified.  This is not
surprising since most of the new stars have come from various infrared surveys,
yielding primarily mid- to late-type T stars.}{These additions bring the number
of known ultracool subdwarfs to around 60.}

\replaced{As a result of Gaia's accomplishments, parallax determinations at optical
wavelengths carried out from the ground targeting individual objects in a
narrow field of view -- such as the NOFS program on the Strand Astrometric
Reflector -- will only continue for a very limited number of stars (e.g.,
certain perturbation candidates).  Ground-based astrometric observations
from large field surveys such as the Large Synoptic Survey Telescope
observations around 2020, will make important contributions for targets
fainter than $G=21$.  Just how accurate the LSST parallaxes will turn out to be
depends on several issues yet to be finalized for LSST, especially
the cadence of the observations, since the LSST determinations will require
correction for differential color refraction and correction from relative to
absolute parallax  values.  Several participants at the 2014 Torino workshop
(cf. de Bruijne 2014) noted that the projected error of the LSST parallaxes
will be $\approx\pm0.4$ mas for stars with SDSS $r$-band magnitudes over the
range from $\sim16.0$ to $\sim 20.0$.  Hence there should be an important
overlap with Gaia's absolute parallax determination.

Extending the Gaia's faint limit to $G=21.0$ -- as discussed at the Torino
workshop -- would provide a smooth transition between Gaia and LSST parallax
determinations as well as increasing the numbers of measured ultracool M-type
dwarfs and subdwarfs by a factor of four.  This should contribute to an
important improvement in our knowledge of the M-type subdwarf luminosity 
function for $M_V \geq14$.}{
{\it Gaia}, a cornerstone mission in the science program of the European Space Agency
\citep[ESA;][]{2016A&A...595A...2G, 2016A&A...595A...1G}, will measure parallaxes
for large numbers of M dwarfs and subdwarfs.  For a {\it Gaia} apparent magnitude of
$G = 20$, the sky-averaged, end-of-mission parallax error is projected to be
$\pm0.3$ mas \citep{2014MmSAI..85..643S}.  Employing the $G$ versus $V$,$V-I$
cubic polynomial relation from \citet[][Table 6]{2010A+A...523A..48J} -- updated
with the coefficients from \citet{2014MmSAI..85..643S} -- we estimate that for a
very-late M-dwarf (typically with $M_V \sim 19$ and $V-I \sim 4.7$;
cf. Figure~\ref{fig:mv}), {\it Gaia} will reach to $V \sim 23.2$, corresponding to a
distance of roughly 70 pc.  Adopting the local luminosity function from
\citet{2015MNRAS.451..149J}, {\it Gaia} should measure parallaxes for
$\sim$ 1700 $M_V \sim 19$ dwarfs, all to an accuracy of 2\% or better.
Similar estimates for the earlier-type (M7 and M8) ultracool dwarfs indicate
that {\it Gaia} will observe them, on average, to distances of 110 pc and 90 pc,
respectively, leading to measurements of roughly 25000 of the M7 \& M8 dwarfs.
Furthermore, the actual {\it Gaia} limiting magnitude will be $G \sim 20.7$, resulting
in measurements for additional stars, with somewhat increased parallax errors.

Thus, {\it Gaia} will provide an orders-of-magnitude greater sample size for M dwarfs,
and, by extension, an increased sample of early-L dwarfs.  For mid-L and
cooler dwarfs, the rapidly fading absolute magnitude in the {\it Gaia} $G$-band
will severely limit the distance at which they will be observed.
Similarly, for ultracool subdwarfs, the {\it Gaia} limiting magnitude and their
low space density will limit the sample size.
Here, the Large Synoptic Survey Telescope
\citep[LSST; ][]{2008SerAJ.176....1I} should extend samples of ultracool
dwarfs to apparent magnitudes much fainter than $G=21$.  Commencing observations
around 2020, LSST will measure parallaxes with an accuracy that will depend on
several issues still to be finalized, especially the cadence of the observations.
(LSST astrometry will require correction for differential color refraction
and correction from relative to absolute parallax values.)
Several participants at the 2014 Torino workshop (cf.
\citealt{2014MmSAI..85..643S}) noted that the projected error of the LSST parallaxes
will be $\approx\pm0.4$ mas for stars with SDSS $r$-band magnitudes over the
range from $\sim16.0$ to $\sim 20.0$.  Hence, there should be an important
overlap with {\it Gaia's} absolute parallax determination.}

\acknowledgements
We sincerely thank Bill van Altena for his continued support of
the NOFS parallax program, especially during the late phases of
the photographic era and the transition to CCD detectors.
Special thanks go to Jim Liebert and Pat Boeshaar for communicating
results from their spectroscopic investigations to us, allowing
our astrometric observations to commence as soon as possible.
Likewise, we thank the members of the 2MASS Rare Objects Team --
especially Davy Kirkpatrick, Neill Reid, and Kelle Cruz -- for
timely alerting us in a timely matter about interesting targets.
\added{We further thank the anonymous referee for many helpful suggestions.} This research has made use of the SIMBAD database, operated at
CDS, Strasburg, France.  We greatly appreciate the dedicated
efforts.
Special thanks are extended to the Engineering Team here at NOFS
-- Mike DiVittorio, Al Rhodes, Mike Schultheis, Alan Bird, and
Steve Sell -- for maintaining the Strand Astrometric Reflector
in top condition.

This publication makes use of data products from the Two Micron All Sky Survey, which is a joint project of the University of Massachusetts and the Infrared Processing and Analysis Center/California Institute of Technology, funded by the National Aeronautics and Space Administration and the National Science Foundation.

This work has made use of data from the European Space Agency (ESA)
mission {\it Gaia} (\url{https://www.cosmos.esa.int/gaia}), processed by
the {\it Gaia} Data Processing and Analysis Consortium (DPAC,
\url{https://www.cosmos.esa.int/web/gaia/dpac/consortium}). Funding
for the DPAC has been provided by national institutions, in particular
the institutions participating in the {\it Gaia} Multilateral Agreement.

\bibliography{dwarf}{}
\facility{USNO:61in (TEK2K, EEV24, TI800), USNO:40in}
\software{DAOPHOT \citep{1987PASP...99..191S}}

\end{document}